\documentclass[journal]{IEEEtran}

\UseRawInputEncoding

\usepackage{cite}
\usepackage{amsmath,amssymb,amsfonts}
\usepackage{mathtools,amssymb,lipsum, nccmath}
\usepackage{graphicx}
\usepackage{cuted}
\usepackage{stfloats}
\usepackage{textcomp}
\usepackage{xcolor}
\usepackage{fancyhdr}
\usepackage{graphicx, nicefrac}
\usepackage{algorithm}
\usepackage{algpseudocode}
\usepackage{multirow}
\usepackage{xfrac}
\usepackage{makecell}
\newcolumntype{?}[1]{!{\vrule width #1}}
\DeclareMathOperator*{\argmin}{arg\,min}
\DeclareMathOperator*{\argmax}{arg\,max}
\allowdisplaybreaks
\usepackage{relsize}
\usepackage{xfrac}
\usepackage[english]{babel}
\usepackage{amsthm,xpatch}
\newtheorem{theorem}{Theorem}
\newtheorem{lemma}[theorem]{Lemma}
\xpatchcmd{\proof}{\hskip\labelsep}{\hskip3\labelsep}{}{} 
\newtheorem{prop}{Proposition}
\usepackage{relsize}
\newcounter{casenum}
\newenvironment{caseof}{\setcounter{casenum}{1}}{\vskip.5\baselineskip}
\newcommand{\case}[2]{\vskip.5\baselineskip\par\noindent {\bfseries Case \arabic{casenum}:} #1\\#2\addtocounter{casenum}{1}}

\usepackage{booktabs}

\algnewcommand\algorithmicforeach{\textbf{for each}}
\algdef{S}[FOR]{ForEach}[1]{\algorithmicforeach\ #1\ \algorithmicdo}
\ifCLASSINFOpdf
 
\fi

\begin{document}

\title{\LARGE
STAR-RIS-Assisted Hybrid NOMA mmWave Communication: Optimization and Performance Analysis}

\author{Muhammad Faraz Ul Abrar, Muhammad Talha,
         Rafay Iqbal Ansari,~\IEEEmembership{Senior Member,~IEEE}, Syed Ali Hassan,~\IEEEmembership{Senior Member,~IEEE,}
        and~Haejoon Jung,~\IEEEmembership{Senior Member,~IEEE}
        
\thanks{M. Faraz, M.Talha and S. A. Hassan are with the School of Electrical Engineering and Computer Science (SEECS), National University of Sciences and
Technology (NUST), 44000, Islamabad, Pakistan (email: \{mabrar.bee17seecs, mtalha.bee17seecs, ali.hassan\}@seecs.edu.pk)}
\thanks{R. I. Ansari is with the Department of Computer and
Information Sciences, Northumbria University Newcastle, UK. (email: rafay.ansari@northumbria.ac.uk)}
\thanks{H. Jung is with the Department of Electronic Engineering, Kyung Hee
University, Yongin 17104, Korea (e-mail: haejoonjung@khu.ac.kr)}
}


\maketitle


\begin{abstract}
Simultaneously reflecting and transmitting reconfigurable intelligent surfaces (STAR-RIS) has recently emerged out as prominent technology that exploits the transmissive property of RIS to mitigate the half-space coverage limitation of conventional RIS operating on millimeter-wave (mmWave). In this paper, we study a downlink STAR-RIS-based multi-user multiple-input single-output (MU-MISO) mmWave hybrid non-orthogonal multiple access (H-NOMA) wireless network, where a sum-rate maximization problem has been formulated. The design of active and passive beamforming vectors, time and power allocation for H-NOMA is a highly coupled non-convex problem. To handle the problem, we propose an optimization framework based on alternating optimization (AO) that iteratively solves active and passive beamforming sub-problems. Channel correlations and channel strength-based techniques have been proposed for a specific case of two-user optimal clustering and decoding order assignment, respectively, for which analytical solutions to joint power and time allocation for H-NOMA have also been derived. Simulation results show that: 1) the proposed framework leveraging H-NOMA outperforms conventional OMA and NOMA to maximize the achievable sum-rate; 2) using the proposed framework, the supported number of clusters for the given design constraints can be increased considerably; 3) through STAR-RIS, the number of elements can be significantly reduced as compared to conventional RIS to ensure a similar quality-of-service (QoS). 
\end{abstract}


\section{Introduction}
The development of various advanced applications such as augmented reality, high quality broadcast demands strict requirements for fifth-generation and beyond (B5G) networks with regards to ultra reliable low latency communications (URLLC) \cite{Intro1}. To achieve such requirements, millimeter-wave (mmWave) technology is of prime interest among researchers and industry, making it a promising technology for driving B5G networks \cite{Intro2}. However, higher sensitivity to blockages due to very short wavelength of mmWave frequency spectrum is a major drawback in achieving such high gains and low latency requirements \cite{Intro3}.

To minimize the effect of channel impairments such as random small-scale fading, path loss and blockages, technologies such as multiple input multiple output (MIMO) and massive MIMO are proposed to be highly beneficial, however, \cite{Intro19,Intro4,Intro5} have shown that the usage of reconfigurable intelligent surfaces (RISs) can significantly enhance energy efficiency (EE) by creating virtual links between base station (BS) and users and achieving higher reflect beamforming gains. RIS can efficiently modify the properties of a large number of passive, low-cost and programmable reflective elements to reflect the incoming signal in the desired direction with the help of a controller \cite{Intro9}. Because of their nearly passive nature, RISs can enhance the communication performance without the need of radio frequency chains in contrast to active relays, and provide low self interference and hardware costs \cite{Intro9}. Moreover, in case of blockage of the direct link between the BS and mobile user, RIS can also be used to realize a smart tunable wireless environment by dynamically modifying the effective blocked end-to-end channel between the BS and users \cite{Intro17}. However, all these studies have considered RIS in reflection mode only. This half-space coverage limits the usage and potential benefits of RIS-assisted communication networks.


Non-orthogonal multiple access (NOMA) is considered to be a key candidate for interference-limited future wireless networks to provide high data rates \cite{Intro18}, whose basic principle is to serve multiple users over the same time-frequency resource block via a common beamforming vector. NOMA has been studied in conjunction with RIS networks to increase coverage range, user fairness and EE of communication networks \cite{Intro8}. The integration of NOMA and RIS can further improve the system performance, as RIS can assist NOMA to dynamically tune the channels of the paired users yielding significant NOMA gains, and likewise, RIS-assisted systems operating on NOMA can achieve much higher SE than OMA. With the aim to overcome the half-space coverage constraint of RIS along with limited possible BS-user association in severe blockage conditions, researchers have recently proposed simultaneously reflecting and transmitting reconfigurable intelligent surfaces (STAR-RIS) to enhance the existing RIS capabilities to serve users omni-directionally, which can transmit and reflect the incoming signal simultaneously by introducing the equivalent surface electric and magnetic currents in the model \cite{Intro15}, thus enabling full-space network coverage around RIS.

A STAR-RIS can operate in three different modes namely, energy splitting (ES), mode switching (MS) and time switching (TS) depending upon use case requirements \cite{mu2021simultaneously}. For the ES mode STAR-RIS, each element may have a different reflection-transmission amplitude coefficient ratio, also known as ES ratio that determines the proportion of signal being reflected or transmitted, respectively. The overall reflection or transmission of signal impinged on STAR-RIS appears as a net effect of the ES ratio of all the elements of the STAR-RIS. Another key feature is that the phases corresponding to reflection and transmission are completely independent, which can be exploited to maximize overall user coverage and performance gains of randomly scattered users.

Although NOMA assists RIS systems in increasing the spectral efficiency (SE) significantly, however, serving all users simultaneously on NOMA via a unified beamformer may not help achieve optimal sum-rate depending upon user-channel correlations. Therefore, we propose an alternative multiple access scheme H-NOMA for RIS and in particular, STAR-RIS systems, in which users are grouped into several clusters. While all the clusters are served using OMA, all the users belonging to a cluster are served simultaneously using NOMA. In the next section, we discuss prior studies related to the conventional RIS and STAR-RIS-based systems.

\subsection{Related Works}
Several works have been done to exploit the benefits of RIS-enabled communication by optimizing joint beamforming, EE and sum-rate maximization in RIS-assisted uplink network\cite{Intro10,Intro11,Intro12,Intro13,Intro14}. Due to the passive nature of RIS, channel estimation of the RIS-assisted communication systems is crucial, and therefore, there have been several studies in \cite{RISChannel1,RISChannel2,RISChannel3,RISChannel4,RISChannel5,refpaper_EnergyEfficiency} to obtain channel state information (CSI) to maximize achievable gains and network coverage. In addition, the authors in \cite{Intro6,Intro7,Intro8} show that RIS-NOMA is a promising technique to achieve higher EE, SE and user fairness compared to OMA-assisted RIS systems for future wireless communication networks. 
However, all of the above-mentioned works consider RIS in reflection mode only, i.e., users are located in only half-coverage space of the RIS.

In contrast, recent studies of the STAR-RIS-empowered communication networks present that by deploying STAR-RIS, the coverage and SE are significantly improved as compared to conventional RIS-assisted system (i.e., with reflection mode only)  \cite{refpaper_spectralefficiency}. The authors in \cite{mu2021simultaneously} propose a power consumption minimization algorithm for ES, TS and MS protocols and prove that TS and ES operating protocols are generally preferable for uni-cast and multi-cast transmissions. Further, \cite{refpaper_sumratemaximization} shows that using NOMA with STAR-RIS can significantly increase the user coverage range as compared to OMA-assisted STAR-RIS system. All these studies have considered either single-user or two-user (one served via reflection and other through transmission of STAR-RIS) system model. Moreover, the authors in \cite{RISChannel6} propose an efficient algorithm for uplink CSI acquisition for the STAR-RIS systems, whereas the work in \cite{RIS_hardware} shows how to implement the STAR-RIS system in practice. A recent work on STAR-RIS in \cite{refpaper_joint_star} solves a similar sum-rate maximization problem, while serving multiple conventional NOMA clusters simultaneously, however, such a choice of multiple access scheme can provide higher gains only when the served users have similar channel conditions.

\subsection{Motivation and Contribution}
Considering its unique features over conventional RIS-assisted systems, a STAR-RIS can enhance the wireless connectivity in $360^\circ$ (i.e., full-space) coverage by introducing the independent transmissive and reflective beams. In order to fully exploit the advantages of STAR-RIS, joint active and passive beamforming optimization needs to be done for multiple users randomly scattered surrounding the STAR-RIS. The literature on STAR-RIS is in its initial phases, and to the best of our knowledge, optimal multiple access scheme and a detailed performance analysis for STAR-RIS systems with a design perspective showing the effect of active, passive beamforming vectors and resource allocation scheme for sum-rate maximization has not been presented yet. This motivates us to develop a complete optimization framework and analyze the impact of each design parameter of a STAR-RIS-assisted system in achieving optimal sum-rate. The main contributions of our work can be summarized as follows:
\begin{itemize}
\item For a multi-cluster hybrid NOMA (H-NOMA)-based STAR-RIS system, we propose optimal user pairing and decoding order strategies using user-channel correlations and channel strengths for two-user NOMA clustering, where each cluster consists of two users (one served via reflection and the other via transmission through the STAR-RIS).   
\item We propose an alternating optimization (AO)-based algorithm to determine the active and passive beamforming vectors, time slot allocation and power allocation. We also show the impact of each of these individual design parameters on the achievable sum-rate.
\item We provide a detailed performance analysis and compare the proposed algorithm with OMA-assisted STAR-RIS and conventional NOMA-assisted STAR-RIS systems. The results show that because of additional degree of freedom in H-NOMA, it outperforms conventional NOMA-based STAR-RIS systems in \cite{refpaper_joint_star}.  
\end{itemize}

\begin{figure}[t!]
\centering
\includegraphics[width=\linewidth]{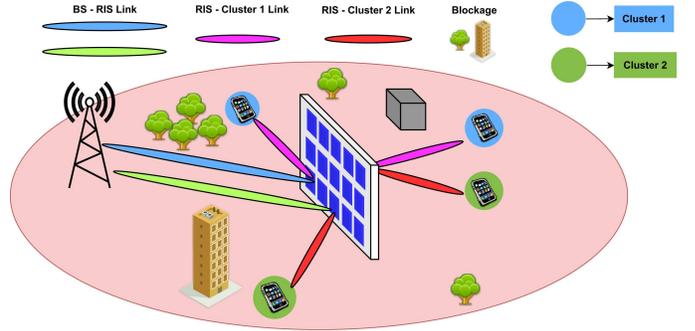}
\caption{A STAR-RIS-assisted H-NOMA system serving four blocked users with $C=2$, $K_c=2$ through reflection and transmission.}
\label{fig:Sysmodel}
\end{figure}

\subsection{Paper Organization and Notation}
The rest of the paper is organized as follows. Section \ref{secSM} presents the system model including signal model, channel model and multiple access schemes. Section \ref{sProbForm} presents the problem formulation and the proposed solution, whereas in Section \ref{sPropSol}, we present optimal solutions to user pairing and decoding order assignment, active and passive beamforming optimizations and joint H-NOMA power coefficient and time-slot allocation problems. Section \ref{sSimRes} presents the performance analysis of the considered STAR-RIS system and other simulation results using the proposed algorithm. Finally, Section \ref{sConclusion} concludes the paper.

Lower-case and upper-case boldface letters denote vectors and matrices, respectively. Further, $\mathbb{C}^{N \times 1}$ denotes a $N$ dimensional complex vector, and $\mathbb{C}^{N \times K}$ represents a $N \times K$ dimensional complex matrix. Also, $(.)^H$ denotes the Hermitian (conjugate transpose), and vec$(.)$ and diag$(.)$ corresponds to vectorization and diagonalization, respectively. $[\mathbf{A}]_{i,j}$ returns the entry of the input matrix corresponding to the $i$-th row and $j$-th column, whereas $\mathbf{A} \circ \mathbf{B}$ represents Hadamard (element-wise) product of the two matrices $\mathbf{A \text{ and } B}$. Also, $|\,.\,|$ corresponds to the magnitude of a complex number or cardinality of the set if the input is a set. In addition, $||\,.\,||_2$ denotes the $\ell_2$-norm, and Real $(.)$, angle$(.)$ returns the real component, and phase of a complex argument, respectively.

\section{System Model}
\label{secSM}
We consider a STAR-RIS-assisted mmWave downlink multi-user multiple-input single-output (MU-MISO) communication scenario with $N_t$ BS antenna elements, $K$ single-antenna users and $M$ STAR-RIS passive units that reflect or transmit the signals coming from the BS towards the intended users. This model is particularly helpful in scenarios where the BS to user link is blocked. Fig. \ref{fig:Sysmodel} shows the considered scenario. The STAR-RIS is assumed to operate in the ES mode, where it is able to both reflect and transmit the incoming BS signal simultaneously to create a link between the BS and the users. In addition, the $K$ supported users are divided into $C$ clusters, where the $c$-th cluster serves a total of $K_c$ users and let $\mathbb{K}_c \triangleq \{1,2, ..., K_c$\} represent the set of all $K_c$ users in cluster $c$, where each user $k$ $\in \,\{r,t \}$ represents the user being served in the reflecting and transmitting mode, respectively. Suppose that $\mathbb{C} \triangleq \{1, 2, ... \,, C\}$ denotes the set of clusters, in which users are grouped and $m \in \mathbb{M} \triangleq \{1, 2, ... , M \}$ denotes the set of STAR-RIS elements. Further, let $T_{\text{max}}$ denote the channel's coherence time, which is divided into $C$ sub-time slots, where each $t_c, \,\forall\, c \in \mathbb{C}$, is the time allocated to serve all users of cluster $c$. Taking into account the practical hardware limitations, discrete STAR-RIS amplitudes and phases are assumed. To be specific, the $m$-th STAR-RIS element's phase corresponds to $\theta^\chi_{m,c} \in \mathbf{\Psi} \triangleq 
\Big\{ 0, \frac{2\pi}{2^{B_1}}, \,.\,.\,. \,^,\, \frac{2\pi(2^{B_1} - 1)}{2^{B_1}} \Big\} $, while its amplitude is denoted by $\beta^\chi_{m,c} \in \mathbf{\Omega} \triangleq\big\{0, \frac{1}{2^{B_2} -1},\, \frac{2}{2^{B_2} -1 }, \,.\,.\,. \, ,\,\frac{2^{B_2} -1 }{2^{B_2} -1}$ \big\}, where $B_1 \text{ and } B_2$ are finite available resolution bits for tuning phase shifts and amplitude coefficients of the RIS elements, respectively, and $\chi \in X \triangleq \{r,t\}$ for the reflection and transmission modes, respectively. For lossless operation, we assume for the $m$-th STAR-RIS element
\begin{equation*}
    \beta^\chi_{m,c} + \beta^{\overline{\chi}}_{m,c} = 1.
\end{equation*}
However, the phase shifts $\theta^\chi_{m,\chi}$ and $\theta^{\overline{\chi}}_{m,c}$ can be independently tuned $\forall \, m \in \mathbb{M}$.

\begin{figure}[t!]
\centering
\includegraphics[width=\linewidth]{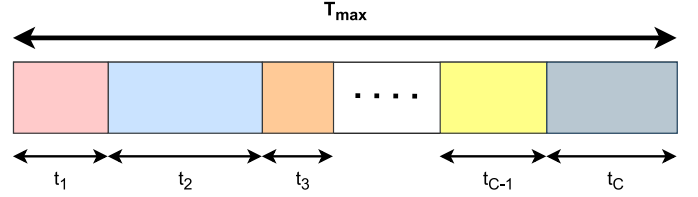}
\caption{Unequal time-slot allocation for different clusters meeting QoS requirement of each within given channel coherence time $T_{\text{max}}$.}
\label{fig:timeslot}
\end{figure}

\subsection{Signal Model}
Let $x_{k,c}$ be the transmitted signal from the BS in time-slot $t_c$ intended for the $k$-th user belonging to the $c$-th cluster, where $\mathbb{E}\Big\{|x_{k,c}|^2\Big\} =1$. The received signal $y_{k,c}$ for the user is 
\begin{equation}
\medmath{y_{k,c} = \underbrace{\mathbf{h}_{k,c}\mathbf{w}_{c}\sqrt{p_{k,c}}\,x_{k,c}}_\text{desired information signal} \,+\, \underbrace{\mathbf{h}_{k,c} \mathbf{w}_{c} \sum_{i\,\in \{\mathbb{K}_c \setminus  \,k\}}\sqrt{p_{i,c}}\,x_{i,c}}_\text{intra-cluster interference} \,+ \, \underbrace{n_{c,k}}_\text{noise}},
\end{equation}
where $p_{k,c}$ represents the allocated power for the $k$-th user of cluster $c$, $\mathbf{w}_c$ is the active beamforming vector associated with the cluster $c$. Also, $\mathbf{h}_{k,c}=\mathbf{ g}_{k,c}^H \mathbf{\Phi}^\chi_c \mathbf{H}$ is the end-to-end channel from the BS to the user, where $\mathbf{g}_{k,c} \in \mathbb{C}^{M \times 1}$ is the channel vector from the STAR-RIS to the $k$-th user of cluster $c$, $\mathbf{H} \in \mathbb{C}^{M \times N_t}$ is the BS-STAR-RIS channel matrix, $\mathbf{\Phi}^\chi_c$ $=$ diag $\left(\begin{bmatrix}\beta^\chi_{1,c} \,e^{j\theta^c_{1,\chi}} ,& \beta^\chi_{2,c}\,e^{j\theta^\chi_{2,c}} , \; ... \; ,& \beta^\chi_{M,c}\,e^{j\theta^\chi_{M,c}}\end{bmatrix}\right) \in \mathbb{C}^{M \times M}$ represents diagonal phase shift matrix, in which $\beta^\chi_{m,c} \in [0,1]$ and $\theta^\chi_{m,c} \in [0,2\pi)$ $\forall \,m\in M$, and $n_{k,c} \sim \mathcal{CN}(0,\,\sigma^2)\,$ is independent and identically distributed (i.i.d) additive white complex Gaussian noise at the receiver with zero mean and $\sigma^2$ variance. We define a function $\boldsymbol{\mathit{state}}(k) \triangleq \chi^{k}$, which returns whether user $k$ is being served through STAR-RIS reflection or transmission. Even though obtaining perfect CSI for both BS-STAR-RIS and STAR-RIS downlink channels is an open research problem, however, the recent attempts in \cite{RISChannel1,RISChannel2,RISChannel3,RISChannel4,RISChannel5} to estimate CSI for conventional RIS-empowered systems with low complexity techniques can also be extended to STAR-RIS-based systems.

\subsection{Channel Model}
We assume that the BS is serving users while operating at mmWave with $(N_t > K)$ antennas; therefore, the BS-STAR-RIS channel matrix is given as
\begin{equation}
  \mathbf{H= H}_\text{LoS} + \mathbf{H}_\text{NLoS} \,,
\end{equation}
 where $\mathbf{H}_\text{LoS}$ and $\mathbf{H}_\text{NLoS}$ represent the line-of-sight (LoS), and non-LoS components of the channel $\mathbf{H}$, respectively. We use the three-dimensional (3D) Saleh-Valenzuela geometry channel model, as in \cite{ChannelRef} for both BS-STAR-RIS and STAR-RIS to user LoS components.Thus, $\mathbf{H}_\text{LoS}$ is given by
 \begin{equation}
\medmath{\mathbf{H}_\text{LoS} = \sqrt{PL(d_{BR})}\Big(\sqrt{N_t M} \; h_\text{LoS}\, \mathbf{h}_r(\theta_\text{LoS}^r,\phi_\text{LoS}^r)\,\mathbf{h}_t^H(\theta_\text{LoS}^t,\phi_\text{LoS}^t )\Big)}.
\end{equation}
\begin{figure*}[b!]
\hrulefill
\begin{equation}
\label{SteerVec1}
\mathbf{h}_t(\theta_l^t,\phi_l^t)=\frac{1}{\sqrt{N_t}} \,\text{vec} \left(\begin{bmatrix}
e^{j\frac{2\pi d}{\lambda} \left((m_y-1) \cos(\theta_l^t)\sin(\phi_l^t) )+ (m_z -1)\sin(\theta_l^t)\right)}\\
\end{bmatrix}_{m_y,m_z}
\right)  \,,\;
\end{equation}
\begin{equation}
\label{SteerVec2}
\mathbf{h}_r(\theta_l^r,\phi_l^r)=\frac{1}{\sqrt{M}} \,\text{vec} \left(\begin{bmatrix}
e^{j\frac{2\pi d}{\lambda} \left((m_y-1) \cos(\theta_l^r)\sin(\phi_l^r) )+ (m_z -1)\sin(\theta_l^r)\right)}\\
\end{bmatrix}_{m_y,m_z}\right). 
\end{equation}
\end{figure*} 
We assume the BS and STAR-RIS to be the rectangular uniform planar arrays (UPAs) with antennas and elements uniformly distributed over $y -z$ plane, i.e., $N_t = N_{ty} \times N_{tz}$, $M = M_y \times M_z$. Also, $h_\text{LoS}$ is the random complex channel gain of the varying channel at a particular time instant, whereas $\mathbf{h}_r(\theta_\text{LoS}^r,\phi_\text{LoS}^r)$ and $\mathbf{h}_t(\theta_\text{LoS}^t,\phi_\text{LoS}^t)$ denote the array steering response vectors at the BS and STAR-RIS, respectively, corresponding to the LoS paths, and are given by \eqref{SteerVec1} and \eqref{SteerVec2}, where $d$ is the inter-element spacing, which is $\frac{\lambda}{2}$ for both BS and STAR-RIS with $\lambda$ being the carrier wavelength. In addition, $\theta_l^t\, \in\begin{bmatrix}-\frac{\pi}{2},\frac{\pi}{2}
\end{bmatrix}\,,\, \phi^t_l \in \begin{bmatrix}
-\frac{\pi}{2},\frac{\pi}{2}
\end{bmatrix}$ are the angles-of-departure (AoD) corresponding to the transmit antennas in the elevation and azimuth plane, respectively, for the $l$-th propagation path. Similarly, $\theta_l^r \in\begin{bmatrix}-\frac{\pi}{2},\frac{\pi}{2}
\end{bmatrix}, \;\phi_l^r \in\begin{bmatrix}-\frac{\pi}{2},\frac{\pi}{2}
\end{bmatrix}  $ are the angles-of-arrival (AoA) of STAR-RIS elements for the elevation and azimuth plane, respectively.

Also, the NLoS component fo the BS-STAR-RIS channel, $\mathbf{H}_\text{NLoS} \sim \mathcal{CN}(0,\,10^{-0.1PL(d_{BR})})$, is modeled as a Rayleigh flat fading channel, where $PL(d_{BR})$ represents the path loss at a distance $d_{BR}$ (the distance from the BS to the STAR-RIS) during signal propagation and, as in \cite{PLRef}, is given by 
\begin{equation}
PL(d) = PL_{do} + 10\eta\log(\frac{d}{d_o}) + \zeta \hspace{6mm} \text{[{dB}]},
\end{equation} 
where $PL_{do}$ is path loss at a reference distance $d_o$, whereas $\eta$ is the path loss exponent. Further, $\zeta$ represents the shadowing of the environment. 
Similarly, the STAR-RIS-user channel model is given by
\begin{equation}
    \mathbf{g} = \mathbf{g}_\text{LoS} + \mathbf{g}_\text{NLoS},
\end{equation}
where $\mathbf{g}_\text{NLoS} \sim \mathcal{CN}(0,\,10^{-0.1PL(d_{RU})})$ is the NLoS channel component of $\mathbf{g}$, while the LoS component $\mathbf{g}_\text{LoS}$ is given by
\begin{equation}
\mathbf{g}_\text{LoS} = \sqrt{PL(d_{RU})}\Big(\sqrt{N_t M} \; g_\text{LoS}\,\mathbf{g}_t(\theta_\text{LoS}^t,\phi_\text{LoS}^t )\Big),
\end{equation}
where $PL(d_{RU})$ is the path loss at the distance $d_{RU}$ (the distance between the STAR-RIS and the user). Also, $g_\text{LoS}$ represents the complex channel gain, whereas $\mathbf{g}_t$ corresponds to the array steering response vector from the STAR-RIS to the users and is given by
\begin{equation}
\mathbf{g}_t(\theta_l^t,\phi_l^t)=\begin{bmatrix}
1 , \;... \;, e^{j\frac{2\pi}{\lambda}(m_y -1)\cos(\theta_l^t)\sin(\phi_l^t)}
\end{bmatrix} 
,
\end{equation}
where $\theta_l^t\;\in \; \big[\frac{\pi}{2},\frac{\pi}{2}\big]$ and  $\phi_l^t \; \in \; \big[-\frac{\pi}{2},\frac{\pi}{2}\big]$.

\subsection{Multiple Access Schemes}
NOMA works on the principle of superposition coding (SC) on the BS to serve multiple users in a cluster and performs successful successive interference cancellation (SIC) on users to remove the intra-cluster-interference (IntraCI). This intrinsic property of NOMA can be exploited to expand the rate region \cite{RateRegion}. Consequently, optimal rates can be achieved if the BS serves the paired users via the STAR-RIS reflection and transmission of the combined signal via an active beamformer $\mathbf{w}_c$ having higher correlation with the paired user's channel vector $\mathbf{h}_{k,c}$ within the same time and frequency resources.

To the best of our knowledge, there is no substantial work on STAR-RIS that explores the choice of multiple access scheme for randomly scattered blocked users in addition to other design parameters. Although NOMA can be used to leverage higher gains through users having correlated channels, most of the practical communication situations might not demonstrate such behaviors, thereby, making NOMA inefficient in such scenarios. On the other hand, STAR-RIS-assisted OMA scheme also limits the achievable sum-rate. For example, time-division multiple access (TDMA) would serve every user in different time-slot. Similarly, frequency-division multiple access (FDMA), in addition to higher resource requirements, would also require higher external hardware control by partitioning the STAR-RIS into various sections in order to serve multiple BS beams simultaneously, where the elements in each section are controlled through their own controller to either reflect or transmit the signals to their associated users. 

Therefore, to acquire a decent trade-off between achievable gains through NOMA and OMA, we pair the users into various clusters, and serve them on TDMA ensuring user quality of service (QoS) requirements. All the users in a cluster are served by the BS simultaneously with NOMA, by allocating different power allocation coefficients, in the time-slot allocated for that cluster, as shown in Fig. \ref{fig:timeslot}. This multiple access scheme has been introduced in \cite{H_NOMA_ref} to maximize the EE of the system, and is referred to as H-NOMA. H-NOMA has been proven an ideal candidate to reduce the NOMA SIC complexity at the receiver of RIS-based systems without having performance degradation because of the increased degree of freedom in terms of the time-slot and power allocation \cite{RIS_HNOMA}. 
The fundamental principles of NOMA, SC and SIC, require the right choice of user decoding order to obtain optimal rates. For single-input-single-output (SISO) systems, NOMA user decoding orders can be optimally decided via users' channel gains only. However, in the considered MU-MISO system, both active beamforming vector and STAR-RIS-user channel can change the user's effective signal-to-noise ratio (SNR), hence for such systems, choice of decoding order becomes an additional challenge, and any of the $K_c!$ order can be used to decode the signals intended for $K_c$ users. 

Let $\vartheta(k_c)$ represent the decoding order of user $k$ in cluster $c$, For convenience, we are dropping the subscript $c$. If $\vartheta(k) = k$ and $\vartheta(i) =i\,$, then the signal intended for each user $i$ will be successively decoded before decoding the $k$-th user signal at user $k$, given that $\left(\vartheta(i) < \vartheta(k)\,\right)\,\forall i \in \mathbb{K}_c$. Let $\vartheta(K_c) \geq \vartheta(k) \geq \vartheta(1),\, \forall\, k \in \mathbb{K}_c$ be the general decoding order for the users of cluster c. Therefore, the achievable signal-to-interference noise ratio (SINR) at user $k$ to decode the $i$-th user signal is given as
\begin{equation}
\label{SINR}
\gamma^c_{k\rightarrow i} = \frac{|\mathbf{h}_{k,c} \mathbf{w}_c|^2\; p_{i,c}} {\sum\limits_{j = i+1}^{K_c} {|\mathbf{h}_{k,c} \mathbf{w}_c|^2 \, p_{j,c}} + \, \sigma^2},
\end{equation}
where $\sum\limits_{j = i+1}^{K_c}\, p_{j,c}$ is the IntraCI, while decoding the $i$-th user's signal at user $k,\, \forall \, \vartheta(k) > \vartheta(i)$, coming from users with higher decoding order than that of user $i$. It is clear that for a given user $k$, whose signal is to be decoded, the IntraCI only depends on the assigned NOMA decoding order to the user $k$ among all the $K_c$ users of the cluster $c$. Thus, the IntraCI for this case can be expressed as
$$\mathcal{I}(k,c) = \mathlarger{\sum}\limits_{j = k+1}^{K_c}\, p_{j,c} = \mathlarger{\sum}_{j \,\ni\, \vartheta(j) > \vartheta(k)} p_{j,c}.$$ Similarly, for decoding the $k$-th user signal at user $k$ the SINR can be given as
\begin{equation}
\label{DecodingSINR}
\begin{split}
\gamma^c_{k\rightarrow k} &= \frac{|\mathbf{h}_{k,c} \mathbf{w}_c|^2\; p_{k,c}} {\sum\limits_{j = k+1}^{K_c} {|\mathbf{h}_{k,c} \mathbf{w}_c|^2 \, p_{j,c}} + \,\sigma^2}, \\
\gamma^c_{k\rightarrow k}&= \frac{|\mathbf{h}_{k,c} \mathbf{w}_c|^2\; p_{k,c}} {\mathcal{I}(k,c) + \,\sigma^2} .
\end{split}
\end{equation}
The corresponding decoding rate can be given as
\begin{equation}
R_{k \rightarrow i }^c = t_c \left(\log_2\left(\,1\, + \, \gamma^c_{k\rightarrow i} \right) \right).
\label{DecodingRate}
\end{equation}

To perform successful SIC to decode signals of all the users of a cluster, power allocation needs to be done carefully to meet QoS requirement of each user while also minimizing the IntraCI. Therefore, the two conditions in \eqref{NOMAconstraints} need to be satisfied.
\begin{figure*}[b]
\hrulefill
\begin{equation}
\label{NOMAconstraints}
\begin{cases}
R_{k \rightarrow i }^c \,\geq\, R_{k-1 \rightarrow i }^c\, \geq\, \,.\, .\, .\, \geq\, R_{i+1 \rightarrow i }^c \,\geq\, R_{i \rightarrow i }^c\,,\quad \quad \forall \,\vartheta(i)\, <\, \vartheta(k)\\
R_{k \rightarrow i }^c \,\geq\, R_{i,c}^\text{min}\,,\,R_{k-1 \rightarrow i }^c \,\geq\, R_{i,c}^\text{min}\,,\, \,.\,.\,.\,,\, R_{i \rightarrow i }^c \,\geq\, R_{i,c}^\text{min}.
\end{cases}
\end{equation}
\end{figure*}

\section{Problem Formulation and Solution}
\label{sProbForm}
We express the problem to maximize the sum-rate for all the users grouped into various clusters in the network. Therefore, we formulate a joint optimization problem consisting of active beamforming, reflection and transmission passive beamforming, user pairing, NOMA decoding order assignment, power allocation coefficients and cluster time-slot allocation. The given problem can be formulated as
\begin{subequations}
\label{ProbForm}
\begin{gather}
\max_{\mathbf{w}_c,\theta_{m,c}^\chi,\beta_{m,c}^\chi,\{\mathbb{K}_c\},\vartheta(k_c),p_k,t_c} \;\sum_{c \,\in \;\mathbb{C}} \,\sum_{k \,\in \,\mathbb{K}_c} R_{k,c}\label{opt1}\\
s.t. \;\|\mathbf{w}_c\|_2^2 \leq P_{max} \,\label{opt2},\\
R^c_{k\rightarrow i} \geq R^\text{min}_{i,c} \;, \,\forall\, i,k\, \in\, \mathbb{K}_c \,, \vartheta(i) \leq \vartheta(k)\, \label{opt3},\\
R^c_{k\rightarrow i} \geq R^c_{i \rightarrow i} \;, \,\forall\, i,k\, \in\, \mathbb{K}_c \,, \vartheta(i) \leq \vartheta(k)\, \label{opt4},\\
\sum_{c \, \in \, \mathbb{C}} t_c = 1,\, t_c \in [0,1]\, \label{opt5},\\
\sum_{k \, \in \, K_c} p_k =  1,\,\, p_k \in [0,1] \,\label{opt6},\\
\sum_{c \, \in\, \mathbb{C}}\; |\mathbb{K}_c| = K\,\label{opt7},\\
1 \leq \vartheta(k) \leq |\mathbb{K}_c| \,,\label{opt8}\\
\beta^\chi_{m,c} + \beta^{\overline{\chi}}_{m, c} = 1\label{opt9},\\
\theta^\chi_{m,c} \,\in \, \mathbf{\Psi} \,,
\beta^\chi_{m,c} \,\in \, \mathbf{\Omega} \,\label{opt10},
\end{gather}
\end{subequations}
\begin{equation*}
\quad \quad \quad \;\, \mathlarger{\forall} \; c \,\in\, \mathbb{C}\,, \;m  \,\in\, \mathbb{M} \,,\, \chi \, \in\,  X\,,
\end{equation*}where $R_{k,c} = R^c_{k\to k}$. The constraint \eqref{opt2} represents the maximum power ($P_{max}$) constraint of transmitting antennas, whereas the constraint \eqref{opt3} refers to the minimum QoS decoding and achievable rate requirement for each user to be served. The constraint \eqref{opt4} refers to SIC constraint following a specific decoding order in a cluster, while \eqref{opt5} refers to the cluster time-slot allocation constraint in terms of percentage of $T_\text{max}$, and \eqref{opt6} represents the NOMA power allocation coefficients constraint for each user of a cluster. Further, \eqref{opt7} refers to user pairing constraint such that each user gets served, and \eqref{opt8} represents the NOMA decoding order assignment constraint. In addition, \eqref{opt9} represents the STAR-RIS transmission and reflection amplitude coefficient constraint, whereas \eqref{opt10} corresponds to the finite resolution bits constraint of STAR-RIS phases and amplitudes. 

The given problem in \eqref{ProbForm} is a highly coupled non-convex optimization problem due to the strong coupling between: 1) NOMA power allocation coefficients, cluster time-slot allocation and STAR-RIS amplitude coefficients 2) active beamformer and the end-to-end BS-user channel due to the STAR-RIS phases-assisted passive beamformer. The non-convexity of the problem lies in the objective function \eqref{opt1} as well as in the constraints \eqref{opt3} and \eqref{opt4}.

\subsection{Proposed Solutions}
\label{sPropSol}
To solve the sum-rate maximization problem, we first perform an optimal user pairing and decoding order scheme, and then approach to solve it by dividing the given problem in multiple sub-problems, i.e, active beamforming, passive beamforming, NOMA power and time-slot allocation optimization problems, and then we solve them independently forming an iterative AO-based algorithm.

\subsubsection{Optimal User Pairing and Decoding Order Scheme}
    We first deal with optimal user pairing of $K$ users into $C$ clusters, and later perform time-slot and power allocation problems. To maximize the sum-rate while also increasing EE, we approach the problem by incorporating two users $\{i,j\}$ in every cluster such that $|\mathbb{K}_c| =2  \,,\,\forall\, c\in \mathbb{C}$ and $\boldsymbol{state}(i) \neq \boldsymbol{state}(j)$. The user pairing with $|\mathbb{K}_c|$ as an optimization parameter has been set apart as a good future work. Therefore, for the scope of this paper, we present a low-complexity user pairing algorithm with even $K$ and $|\mathbb{K}_c| =2$ that presents a near optimal solution.  Let $\mathbf{Q}$ be a $K \times K$ pairing symmetric matrix such that
\begin{equation}
\label{userpair}
\begin{cases}
[\mathbf{Q}]_{i,j}= \psi, \quad \quad \quad \forall \,i \neq j \text{ and } \boldsymbol{\mathit{state}}(i) \neq \boldsymbol{\mathit{state}}(j)\\
[\mathbf{Q}]_{i,j}=0, \quad\, \quad \quad \forall \,i = j$ \text{or} $\boldsymbol{\mathit{state}}(i) = \boldsymbol{\mathit{state}}(j) \\
\sum\limits_{i=1}^K [\mathbf{Q}]_{i,j}=1, \quad \quad \forall\, j \in [1,K]\\
\sum\limits_{j=1}^K [\mathbf{Q}]_{i,j}=1, \quad \quad \forall\, i \in [1,K]
\end{cases}
\end{equation}
where $\psi \in \, \{1,0\}$ represents paired and non-paired users, respectively. Obviously $[\mathbf{Q}]_{i,i} = [\mathbf{Q}]_{j,j} = 0\,$ corresponds to self-pairing. Then the given sub-problem can be modeled as
\begin{subequations}
\label{userPairProb}
\begin{gather}
\label{userpair1}
\max_{\mathbf{Q}}  \sum_{i=1}^K \,\sum_{j=1}^K \,[\mathbf{Q}]_{i,j}\,R_{i,j}, \\
\text{subject to}\,\, \eqref{userpair},
\end{gather}
\end{subequations}
where $R_{i,j}$ is given by \eqref{DecodingRate}.
We handle the above user pairing sub-problem by segregating users in two groups $\mathbf{B}_1$ and $\mathbf{B}_2$ of equal size ($\frac{K}{2}$), on the basis of their channel strength $\|\mathbf{h}_k\|^2 \; \forall \,k \in [1,K]$, while maximizing channel correlations with feasible users consequently maximizing the sum-rate. 

We also define users' channel correlation and users' state matrix $\mathbf{C} \text{ and } \mathbf{S}$, respectively, such that $\mathbf{[C]}_{i,j}= |\mathbf{h}_j\mathbf{h}^H_i|^2$ where $\mathbf{h}_i \text{ and }\mathbf{h}_j $ corresponds to channels of group $\mathbf{B}_1 \text{ and } \mathbf{B}_2$, respectively. For the user state matrix, let $[\mathbf{S}]_{i,j} = 1$, if users $i\,,\,j$ have different states allowing a feasible pair, otherwise 0. The details of solving \eqref{userpair} is summarized in \textbf{Algorithm \ref{Alg:UserPair}} which transforms the original problem. Therefore, we proceed the problem by utilizing the obtained $\mathbf{C}$ and solve the original problem as an assignment linear programming problem (LPP) given by
\begin{subequations}
\label{UserPairProb2}
\begin{gather}
\max_{\mathbf{A}}  \sum_{i=1}^{\nicefrac{K}{2}} \,\sum_{j=1}^{\nicefrac{K}{2}}  \,\mathbf{[C]}_{i,j}\,\mathbf{[A]}_{i,j} \label{ALPP1} \\
s.t.\,\, \sum_{i=1}^{\nicefrac{K}{2}} {\mathbf{[A]}_{i,j}}=1, \quad \forall\, j=\{1,2,\,...\,,\frac{K}{2}\}, \label{ALPP2}\\
\sum_{j=1}^{\nicefrac{K}{2}} {\mathbf{[A]}_{i,j}}=1, \quad \forall\, i=\{1,2,\,...\,,\frac{K}{2}\},\label{ALPP3}\\
\mathbf{[A]}_{i,j} \in \{0,1\} \label{ALP41}.
\end{gather}
\end{subequations}

This LPP is a convex problem that can be easily solved by optimization solvers such as CVX, or Hungarian algorithm \cite{HungarianAlgo} to obtain $\mathbf{A}^\star$, which can be utilized to obtain $\mathbb{K}_c^\star$. The given algorithm performs user-pairing via channel correlation, however at a later stage, we present \textbf{Algorithm \ref{Alg:PropAlg}}, which optimizes the active beamformer such that $ |\mathbf{h}_j\mathbf{h}_i^H|^2 \rightarrow n(|\mathbf{h}_j\mathbf{w}_c|^2)$, where $n$ is a scaling constant. The proposed algorithm provides the optimal solution for the sum-rate maximization objective function for the considered case.

The optimal strategy to allocate NOMA user decoding order $\vartheta(k)$ for maximum sum-rate is to assign higher decoding order to the user with higher channel strength $\|\mathbf{h}_k\|^2$. Therefore, optimal decoding order is assigned such that $\vartheta(i) > \vartheta (j)\, \forall \, i \, \in \,\mathbf{B}_1, \, j\, \in \, \mathbf{B}_2$. 
The feasible user pairing matrix $\mathbf{A}$ is a $\frac{K}{2} \times \frac{K}{2}$ Boolean matrix representing a feasible NOMA pair among two users, and hence it can easily be shown that by satisfying modified constraints \eqref{ALPP2} and \eqref{ALPP3}, the original constraint in \eqref{opt7} is satisfied. Similarly, the decoding order assignment in \eqref{UserPairProb2} satisfies \eqref{opt8}. The computational complexity of \textbf{Algorithm 1} is governed primarily by solving \eqref{UserPairProb2}, which is a function of total supported users $K$, therefore, the overall complexity is given by $O (KN_t + (\frac{K}{2})^3) $.

 \begin{algorithm}[t!]
 \caption{User Pairing and Decoding Order Algorithm for \eqref{userPairProb}}
 \label{Alg:UserPair}
 \vspace{1mm}
 \hspace*{\algorithmicindent} \textbf{Input} : $\{\mathbf{h}_k\},K$ \\
 \hspace*{\algorithmicindent} \textbf{Output} : $\{\mathbb{K}_c^\star\},\, \{\vartheta(k_c)\}$
 \begin{algorithmic}[1]
  \State Initialize $\mathbf{\Tilde{h}} : \{\mathbf{\|h}_k\|\} \,\forall \,k \in [1,K]$
  \State Update $\mathbf{\Tilde{h}} =$ \textbf{SORT}($\mathbf{\Tilde{h}}$) $\ni \Tilde{h}_1 \geq \Tilde{h}_2 \geq ... \geq \Tilde{h}_K$
  \State Set $\mathbf{B}_1 = \{i\},\, \mathbf{B}_2 = \{j\}\;\forall\, i \in \left\{1,2,3,...,\frac{K}{2}\right\},$ and $j \in \left\{\frac{K}{2}+1,...,K\right\}$ from $\mathbf{\Tilde{h}}$
  \State $\mathbf{[C]}_{i,j} \gets \{|\mathbf{h}_j\mathbf{h}_i^H|^2\}\,\forall \, i \in \mathbf{B}_1, \,j \in \mathbf{B}_2$ 
\State $\mathbf{[S]}_{i,j} \gets \{ \mathit{\boldsymbol{state(i)}} \,\boldsymbol{\oplus}\, \mathit{\boldsymbol{state(j)}}\} \,\forall\, i \in \mathbf{B}_1, \,j \in \mathbf{B}_2$
  \State Update $\mathbf{C} \gets \mathbf{C} \circ \mathbf{S}$
  \State Solve \eqref{UserPairProb2} to obtain $\mathbf{A^\star}$ 
\State Pair into $\mathbb{K}_c^\star \gets \{i,j\}\, \forall\, i,j \ni \mathbf{[A^\star]}_{i,j} =1,\text{where } \forall \,c \in \mathbb{C}$
\State Assign $\vartheta(i) \gets |\mathbb{K}_c|\,;\, \vartheta(j) \gets |\mathbb{K}_c| - 1  \iff i \in \mathbf{B}_1, \, j \in \mathbf{B}_2$
  \\
  \Return {}$\{\mathbb{K}_c^\star\},\,\{\vartheta(k_c)\}$
 \end{algorithmic} 
 \end{algorithm}

To optimize the remaining design parameters, the problem in \eqref{ProbForm} has been handled by solving the following three sub-problems given as

\begin{subequations}
\label{pabf}
\begin{align}
\max_{\mathbf{w}_c}  \sum_{c \,\in \;\mathbb{C}} \,\sum_{k \,\in \,\mathbb{K}_c} R_{k,c},\label{subp1}\\
s.t. \; \eqref{opt2},\eqref{opt3}, \eqref{opt4}.
\end{align}
\end{subequations}
\begin{subequations}
\begin{align}
\max_{\theta_{m,c}^\chi, \beta_{m,c}^\chi}  \sum_{c \,\in \;\mathbb{C}} \,\sum_{k \,\in \,\mathbb{K}_c} R_{k,c},\label{subp2}\\
s.t. \; \eqref{opt3},\eqref{opt4},\eqref{opt9},\eqref{opt10}.
\end{align}
\end{subequations}
\begin{subequations}
\begin{align}
\max_{p_k,t_c}  \sum_{c \,\in \;\mathbb{C}} \,\sum_{k \,\in \,\mathbb{K}_c} R_{k,c},\label{subp3}\\
s.t. \; \eqref{opt3},\eqref{opt4},\eqref{opt5},\eqref{opt6}.
\end{align}
\end{subequations}
where these problems correspond to active beamforming optimization, passive beamforming (STAR-RIS) optimization and the joint power and time-slot allocation optimization, respectively. 

\subsubsection{Active Beamforming Optimization}
Since the objective function is non-convex with respect to $\mathbf{w}_c$, we modify the original sub-problem. Suppose $\mathbf{s}_{k,c} \triangleq \mathbf{g}_{k,c}^H \mathbf{\Phi}^c_\chi \mathbf{H} $ , $\mathbf{J}_{k,c} \triangleq \mathbf{s}_{k,c}^H \mathbf{s}_{k,c}$ and $\mathbf{\Upsilon}_{k,c} \triangleq \{\gamma_{k,c}\} \; \forall \,k \in \mathbb{K}_c, \, c \in \mathbb{C}$, are the set of auxiliary variables. Let $r^\text{min}_{k,c} = 2^{R^\text{min}_{k,c}} -1$, be the SNR corresponding to the minimum QoS rate requirement $R^\text{min}_{k,c}$. The constraint \eqref{opt3} can also be written as

\begin{gather*}
r^\text{min}_{k,c} \leq \gamma^c_{k\rightarrow i}\\
r^\text{min}_{k,c} \leq \frac{|\mathbf{h}_{k,c} \mathbf{w}_c|^2\; p_{i,c}} {\sum\limits_{j = i+1}^{K_c} {|\mathbf{h}_{k,c} \mathbf{w}_c|^2 \, p_{j,c}} + \, \sigma^2}\\
r^\text{min}_{k,c}\left({\sum\limits_{j = i+1}^{K_c} {|\mathbf{h}_{k,c} \mathbf{w}_c|^2 \, p_{j,c}} + \, \sigma^2} \right) \leq |\mathbf{h}_{k,c} \mathbf{w}_c|^2\; p_{i,c}\\
r^\text{min}_{k,c} \,\sigma^2 \leq |\mathbf{h}_{k,c} \mathbf{w}_c|^2 \,\rho_{i,c}\;,
\text{where}\quad \rho_{i,c} = p_{i,c} - \,r_{k,c}^\text{min} \sum\limits_{j = i+1}^{K_c} {p_{j,c}}.
\end{gather*}
Consequently, the problem \eqref{subp1} can be re-written as  
\begin{subequations}
\label{abfconv}
\begin{gather}
\max_{\mathbf{w}_{k,c},\gamma_{k,c}} \, \sum\limits_{c \,\in \,\mathbb{C}} \sum\limits_{k \,\in \,\mathbb{K}_c} \,t_c\left(\log_2\,\left(1 + \gamma_{k,c}\right)\right) \label{abf1},\\
s.t. \;| \mathbf{s}_{k,c}\,\mathbf{w}_{c}|^2 \rho_{k,c} \geq r^\text{min}_{k,c} \,\sigma^2 \,,\label{abf2}\\
\frac{| \mathbf{s}_{k,c}\,\mathbf{w}_{c}|^2\; p_{k,c}}{\sum\limits_{j = k+1}^{K_c} |\mathbf{s}_{k,c} \mathbf{w}_c|^2 \, p_{j,c} + \, \sigma^2} \geq \gamma_{k,c}\,,\label{abf3}\\
\eqref{opt2} .
\end{gather}

\end{subequations}

The modified active beamforming sub-problem in \eqref{abfconv}, although having convex objective function, is still non-convex due to constraints \eqref{abf2} and \eqref{abf3}. The sub-problem is further modified by defining auxiliary variables $\{\nu_{k,c}\} \,\forall\, k \in \mathbb{K}_c\,,\, c \in \mathbb{C}$. Applying successive convex approximation (SCA) method using first-order Taylor series for the non-convex term in \eqref{abf2} around local point $\mathbf{\widetilde{w}}_{c}$, it can be further re-written as 
\begin{equation}
\begin{split}
 |\mathbf{s}_{k,c}\mathbf{w}_{c}|^2 &\geq |\mathbf{s}_{k,c}\mathbf{\widetilde{w}}_{c}|^2 + 2\,\text{{Re}}\,\big(\big(\mathbf{\widetilde{w}}^H_{c}\mathbf{J}_{k,c}) \,(\mathbf{w}_{c} - \mathbf{\widetilde{w}}_{c}\big)\big)\; \text{or}\,\\
|\mathbf{s}_{k,c}\mathbf{w}_{c}|^2 &\geq 2\,\text{{Re}}\,\big(\mathbf{\widetilde{w}}^H_{c}\mathbf{J}_{k,c}\mathbf{w}_{c}) -  |\mathbf{s}_{k,c}\mathbf{\widetilde{w}}_{c}|^2\\
&\geq \tau_{k,c}\,(\mathbf{w}_c).
\end{split}
\end{equation}
Also, \eqref{abf3} can be split into two sub-constraints given by
\begin{align}
\frac{| \mathbf{s}_{k,c}\,\mathbf{w}_c|^2\; p_{k,c}}{\nu_{k,c}} \geq \gamma_{k,c} \label{abf4},\\
 \sum\limits_{j = k+1}^{K_c} |\mathbf{s}_{k,c} \mathbf{w}_c|^2 \, p_{j,c} + \, \sigma^2 \leq \nu_{k,c}. \label{abf5}
\end{align}

The constraint in \eqref{abf4} is also non-convex which is again written using first-order Taylor series for $\mathbf{w}_c$ and ${\nu_{k,c}}$ around $\mathbf{\widetilde{w}}_{k,c}$ and $\widetilde{\nu}_{k,c}$, respectively, as

\begin{equation}
\begin{split}
\frac{| \mathbf{s}_{k,c}\,\mathbf{w}_c|^2\;}{\nu_{k,c}} &\geq \frac{2\,\text{Re}\big(\mathbf{\widetilde{w}}_{c}^H\mathbf{J}_{k,c}\mathbf{w}_{c}\big)}{\widetilde{\nu}_{k,c}} - \left(\frac{|\mathbf{s}_{k,c}\mathbf{\widetilde{w}}_c|}{\widetilde{\nu}_{k,c}} \right)^2 \nu_{k,c}\\
&\geq \tau_{k,c}\,(\mathbf{w}_c,\nu_{k,c}).
\end{split}
\end{equation}
Using the modified constraints, the modified active beamforming optimization problem is given as
\begin{subequations}
\label{ABF}
\begin{gather}
\max_{\mathbf{w}_{k,c},\gamma_{k,c}} \, \sum\limits_{c \,\in \,\mathbb{C}} \sum\limits_{k \,\in \,\mathbb{K}_c} \,t_c\left(\log_2\,\left(1 + \gamma_{k,c}\right)\right),\\
s.t. \;\tau_{k,c}\,(\mathbf{w}_c) \,\rho_{k,c} \geq r^\text{min}_{k,c} \,\sigma^2 ,\\
\tau_{k,c}\,(\mathbf{w}_c,\nu_{k,c}) \,p_{k,c} \geq \gamma_{k,c},\\
\eqref{abf5}\,,\eqref{opt2}.
\end{gather}
\end{subequations}

Finally, the problem in \eqref{ABF} is a convex problem which provides a locally optimal solution around points $\mathbf{\widetilde{w}}_{k,c}$ and $\widetilde{\nu}_{k,c}$. It is noted that the solution provided by \eqref{ABF} will be a lower-bound of the optimal solution to the original problem in \eqref{pabf}.

\subsubsection{Reflection and Transmission Passive Beamforming Optimization}
To tackle the highly coupled STAR-RIS unit's reflection as well as the transmission phases and amplitudes, we take the sub-problem in \eqref{subp2} and modify the original objective function by defining $\mathbf{u}^\chi_c \triangleq$ 
$\begin{bmatrix}
\beta^\chi_{1,c}\,e^{j\theta^\chi_{1,c}} \;,& \beta^\chi_{2,c}\,e^{j\theta^\chi_{2,c}} \;,\, .\,.\,. \,,\; \beta^\chi_{M,c}\,e^{j\theta^\chi_{M,c}}
\end{bmatrix} \; \forall\, c \in \mathbb{C}\;$
, where for convenience of calculation, $\beta^\chi_{m,c} \in \left[0,1\right] \text{ and } \theta^\chi_{m,c} \in \left[0,2\pi\right) \,,\,\forall\,
m \in \mathbb{M}$, i.e. , we relax the constraint \eqref{opt10} temporarily by initially finding continuous valued optimal phases and amplitudes, and later on, find their discrete equivalents. We define the end-to-end channel $\mathbf{h}_{k,c}$ in terms of $\mathbf{u}^\chi_c$ as, $\mathbf{\overline{h}}_{k,c} \triangleq \text{ diag}\left(\mathbf{g}^H_{k,c}\right)\mathbf{H w}_{c}$, therefore, the SINR in \eqref{SINR} can also be expressed as
\begin{equation}
\label{PBFsnr}
\text{SINR}=\frac{|\mathbf{u}_c^\chi\, \mathbf{\overline{h}}_{k,c}|^2 \,p_{k,c}}{\sum\limits_{j = i+1}^{K_c} |\mathbf{u}_c^\chi\, \mathbf{\overline{h}}_{k,c}|^2 \, p_{j,c} + \, \sigma^2}.
\end{equation}
The problem \eqref{subp2} is modified in terms of SINR in \eqref{PBFsnr} by defining set of auxiliary variables $\{\gamma_{k,c}\}$ and can be written as

\begin{subequations}
\label{pbfsub}
\begin{gather}
\max_{\mathbf{u}^\chi_c,\gamma_{k,c}} \, \sum\limits_{c \,\in \,\mathbb{C}} \sum\limits_{k \,\in \,\mathbb{K}_c} \,t_c\left(\log_2\,\left(1 + \gamma_{k,c}\right)\right),\\
s.t. \;|\mathbf{u}_c^\chi    \,\mathbf{\overline{h}}_{k,c}|^2 \,\rho_{k,c} \geq r_{k,c}^\text{min}\,\sigma^2 \label{pbf1},
\\
\frac{|\mathbf{u}_c^\chi\, \mathbf{\overline{h}}_{k,c}|^2 \,p_{k,c}}{\sum\limits_{j = i+1}^{K_c} |\mathbf{u}_c^\chi \, \mathbf{\overline{h}}_{k,c}|^2 \, p_{j,c} + \, \sigma^2} \geq \gamma_{k,c}. \label{pbf2}
\end{gather}
\end{subequations}

The problem in \eqref{pbfsub} is still non-convex because of the non-convexity of the constraints \eqref{pbf1} and \eqref{pbf2}. Auxiliary variables $\mathbf{Q}_{k,c} \triangleq \mathbf{\overline{h}}_{k,c}\, \mathbf{\overline{h}}^H_{k,c}$ $ \text{and}\,\{\upsilon_{k,c}\}$ are introduced and the constraint \eqref{pbf2} is split into two sub-constraints as
\begin{gather}
\frac{|\mathbf{u}_c^\chi\, \mathbf{\overline{h}}_{k,c}|^2 \,p_{k,c}}{\upsilon_{k,c}} \geq \gamma_{k,c} \label{pbf3},\\
\sum\limits_{j = k +1}^{K_c}|\mathbf{u}_c^\chi\, \, \mathbf{\overline{h}}_{k,c}|^2 \,p_{j,c} \, + \, \sigma^2  \leq \upsilon_{k,c}. \label{pbf4}
\end{gather}

Now, using a similar approach as done previously, we perform SCA assisting first-order Taylor approximations for left hand sides of the constraints \eqref{pbf1} and \eqref{pbf3} for $\mathbf{u}_c^\chi$ and $\mathbf{\upsilon}_c^\chi$ around $\mathbf{\widetilde{u}}_{k,c}$ and $\mathbf{\widetilde{\upsilon}}_{k,c}$ given by

\begin{gather}
|\mathbf{u}_c^\chi \,\mathbf{\overline{h}}_{k,c}|^2 \geq 2 \,\text{Re} \left(\mathbf{\widetilde{u}}_c^\chi \mathbf{Q}_{k,c} (\mathbf{u}_c^\chi)^H \right) -\, |\mathbf{\widetilde{u}}_c^\chi \mathbf{\overline{h}}_{k,c}|^2 = \tau_{k,c}\,(\mathbf{u}_c^\chi) \label{pbf5}\\
\medmath{\frac{|\mathbf{u}_c^\chi\, \mathbf{\overline{h}}_{k,c}|^2}{\upsilon_{k,c}} \geq \frac{2\,\text{Re} \left( \mathbf{\widetilde{u}}_c^\chi \mathbf{Q}_{k,c} (\mathbf{u}_c^\chi)^H \right)}{\widetilde{\upsilon}_{k,c}} - \left(\frac{|\mathbf{\widetilde{u}}_c^\chi \mathbf{\overline{h}}_{k,c}|}{\widetilde{\upsilon}_{k,c}}\right)^2 = \tau_{k,c}\, (\mathbf{u}_c^\chi,\upsilon_{k,c})}. \label{pbf6}
\end{gather}

Finally, the problem in \eqref{pbfsub} can be expressed using \eqref{pbf5} and \eqref{pbf6} as 
\begin{subequations}
\label{pbfsubp2}
\begin{gather}
\max_{\mathbf{u}_c^\chi,\gamma_{k,c}} \, \sum\limits_{c \,\in \,\mathbb{C}} \sum\limits_{k \,\in \,\mathbb{K}_c} \,t_c\left(\log_2\,\left(1 + \gamma_{k,c}\right)\right),\\
s.t. \;\tau_{k,c} \, (\mathbf{u}_c^\chi)\, \rho_{k,c} \geq r^\text{min}_{k,c} \,\sigma^2, \\
\;\tau_{k,c} \, (\mathbf{u}_c^\chi, \upsilon_{k,c})\, p_{k,c} \geq \gamma_{k,c},\\
\eqref{pbf4}.
\end{gather}
\end{subequations}

The convex problem in \eqref{pbfsubp2} is used to obtain continuous $\mathbf{u}_c^\chi$, and then the feasible solution ${\mathbf{u}_c^\chi}'$ satisfying \eqref{opt8} is obtained by a mean squared error (MSE) mapper (e.g., binary searched-based) as
\begin{subequations}
\label{pbfdiscrete1}
\begin{align}
{\theta_m^\chi}' &= \argmin_{\theta\, \in\, \mathbf{\Psi}}|\,\theta - \text{angle} (\mathbf{u}_c^\chi)\,|^2 ,\\
{\beta_m^\chi}' &= \argmin_{\beta\,\in\, \mathbf{\Omega}} |\,\beta - |\mathbf{u}_c^\chi|\,|^2.
\end{align}
\end{subequations}
Finally, the ${\mathbf{u}_c^\chi}'$ is obtained using \eqref{pbfdiscrete1} as
\begin{equation}
\label{pbfdiscrete2}
{\mathbf{u}_c^\chi}' =
\begin{bmatrix}
{\beta^\chi_{1,c}}'\,e^{j{\theta^\chi_{1,c}}'}\,,& {\beta^\chi_{2,c}}'\,e^{j{\theta^\chi_{2,c}}'} \,,\, .\,.\,. \,,\, {\beta^\chi_{M,c}}'\,e^{j{\theta^\chi_{M,c}}'}
\end{bmatrix}.
\end{equation}

\subsubsection{Power and Time-slot Allocation}
The original problem of the power allocation is non-convex because of the objective function, and to maximize the system sum-rate both time-slot allocation and power allocation should be jointly optimized, as the following sub-problem:
\begin{subequations}
\begin{gather}
\max_{t_c,p_{k,c}} \, \sum\limits_{c \,\in \,\mathbb{C}} \sum\limits_{k \,\in \,\mathbb{K}_c} \,t_c\left(\log_2\,\left(1 + \gamma_{k,c}\right)\right), \\
s.t.\; \eqref{opt3},\eqref{opt4},\eqref{opt5},\eqref{opt6}.
\end{gather}
\end{subequations}

Therefore, we derive analytical solutions for H-NOMA parameters such as user power allocation coefficients and cluster time-slot. First, optimal time-slot allocation and user power allocation are performed for all clusters except one with the highest aggregated user channel strength $\mathlarger{\sum\limits_{k \,\in\, \mathbb{K}_c} \|\mathbf{h}_{k,c}\|^2}$, and then the remaining cluster's optimal allocations are handled separately. We solve a system of equations for all such clusters $ c \in \{\mathbb{C} \setminus C\}$ in a way that their each user $k$ is served at $R^\text{min}_{k,c}$, which are given by 
\begin{equation}
\label{RminForEachUser1}
\begin{gathered}
\medmath{t_c\left(\log_2 \left(1 + \frac{|\mathbf{h}_1 \mathbf{w}_c|^2\; p_{1,c}} {\sum\limits_{j = 1}^{K_c} {|\mathbf{h}_{1} \mathbf{w}_c|^2 \, p_{j,c}} + \,\sigma^2}\right) \right)= R^\text{min}_{1,c}},\\
\medmath{t_c\left(\log_2 \left(1 + \frac{|\mathbf{h}_2 \mathbf{w}_c|^2\; p_{2,c}} {\sum\limits_{j = 2}^{K_c} {|\mathbf{h}_{2} \mathbf{w}_c|^2 \, p_{j,c}} + \,\sigma^2}\right) \right)= R^\text{min}_{2,c}},\\
\setbox0\hbox{=}\mathrel{\makebox[\wd0]{\hfil\vdots\hfil}} \\
\medmath{t_c\left(\log_2 \left(1 + \frac{|\mathbf{h}_{K_c} \mathbf{w}_c|^2\; p_{K_c}} {\sigma^2}\right) \right)= R^\text{min}_{K_c}\;, c \neq C}.
\end{gathered}
\end{equation} 
Also, the sum-rate computed based on \eqref{RminForEachUser1} can be given as
\begin{equation}
\label{RminForEachUser2}
\medmath{t_c\left(\sum\limits_{k\,\in\,\mathbb{K}_c}\log_2 \left(1 + \frac{|\mathbf{h}_{k,c} \mathbf{w}_c|^2\; p_{k,c}} {\sum\limits_{j = k+1}^{K_c} {|\mathbf{h}_{k,c} \mathbf{w}_c|^2 \, p_{j,c}} + \,\sigma^2}\right)\right) = \sum\limits_{k=1}^{K_c} {R^\text{min}_{k,c}}}.
\end{equation}

For two users $\{i,j\} \in \mathbb{K}_c\,,\forall\,c \in \{\mathbb{C} \setminus C\}$ such that $\vartheta(i) < \vartheta(j)$, the following cluster sum-rate should therefore be satisfied 
\begin{align*}
&t_c\left(\log_2 \left(1 + \frac{|\mathbf{h}_i \mathbf{w}_c|^2\; p_{i,c}} {p_{j,c}|\mathbf{h}_i\mathbf{w_c}|^2 + \sigma^2}\right) + \log_2 \left(1 + \frac{|\mathbf{h}_j \mathbf{w}_c|^2\; p_{j,c}} {\sigma^2}\right) \right) 
\nonumber\\
&= R^\text{min}_{i,c} + R^\text{min}_{j,c}.
\end{align*}

The following systems of equations are solved to obtain optimal time-slot and power allocation
\begin{equation}
\label{time_pow_initial}
\begin{cases}
\begin{aligned}
t_c\left(\log_2 \left(1 + \frac{|\mathbf{h}_i \mathbf{w}_c|^2\; p_{i,c}} {p_{j,c}|\mathbf{h}_i\mathbf{w_c}|^2 + \sigma^2}\right) \right)= R^\text{min}_{i,c},\\
t_c\left(\log_2 \left(1 + \frac{|\mathbf{h}_j \mathbf{w}_c|^2\; p_{j,c}} {\sigma^2}\right) \right) =R^\text{min}_{j,c}.
\end{aligned}
\end{cases}
\end{equation}

\begin{caseof}
\case{Paired users having similar QoS requirements
($R^\text{min}_{i,c} \approx  R^\text{min}_{j,c})$}
Since the QoS requirements of the paired users $R^\text{min}_{i,c},\,R^\text{min}_{j,c}$ are not different significantly, the closed-form solution to the power and time-slot allocations can be derived as
\begin{gather*}
t_c = \frac{R^\text{min}_{j,c}}{
\log_2 \left(1 + \cfrac{|\mathbf{h}_j \mathbf{w}_c|^2\; p_{j,c}} {\sigma^2}\right) 
} 
\\
\medmath{
\frac{R^\text{min}_{j,c}}{
\log_2 \left(1 + \frac{|\mathbf{h}_j \mathbf{w}_c|^2\; p_{j,c}} {\sigma^2}\right) 
}
\log_2 \left(1 + \frac{|\mathbf{h}_i \mathbf{w}_c|^2\; \left(1 - p_{j,c}\right)} {p_{j,c}|\mathbf{h}_i\mathbf{w_c}|^2 + \sigma^2}\right) 
= R^\text{min}_{i,c}}.
 \end{gather*}

Since $\mathlarger{\left(\nicefrac{R^\text{min}_{j,c}}{R^\text{min}_{i,c}}\right)} \to 1\,,$ we approximate the QoS requirements of the two users by defining $R^\text{min}_{k,c} \triangleq R^\text{min}_{i,c} \approx  R^\text{min}_{j,c}$ 
 \begin{gather*}
\medmath{\log_2 \left(1 + \frac{|\mathbf{h}_i \mathbf{w}_c|^2\; \left(1 - p_{j,c}\right)} {p_{j,c}|\mathbf{h}_i\mathbf{w_c}|^2 + \sigma^2}\right) = \log_2 \left(1 + \frac{|\mathbf{h}_j \mathbf{w}_c|^2\; p_{j,c}} {\sigma^2}\right)} \\
\frac{|\mathbf{h}_i \mathbf{w}_c|^2\; \left(1 - p_{j,c}\right)} {p_{j,c}|\mathbf{h}_i\mathbf{w_c}|^2 + \sigma^2} = \frac{|\mathbf{h}_j \mathbf{w}_c|^2\; p_{j,c}} {\sigma^2} \\
\medmath{p_{j,c}^2\left( |\mathbf{h}_i\mathbf{w}_c|^2 \,|\mathbf{h}_j\mathbf{w}_c|^2|\right) + p_{j,c} |\mathbf{h}_j\mathbf{w}_c|^2 \sigma^2 = 
|\mathbf{h}_i\mathbf{w}_c|^2 \sigma^2 - p_{j,c} |\mathbf{h}_i\mathbf{w}_c|^2 \sigma^2} \end{gather*}

\begin{gather}
\label{timepowQuad}
\medmath{p_{j,c}^2\left( |\mathbf{h}_i\mathbf{w}_c|^2\,|\mathbf{h}_j\mathbf{w}_c|^2|\right) + p_{j,c} \,\sigma^2 \,( |\mathbf{h}_i\mathbf{w}_c|^2 + |\mathbf{h}_j\mathbf{w}_c|^2 ) = \sigma^2\,
 |\mathbf{h}_i\mathbf{w}_c|^2}.
\end{gather}
The equation in \eqref{timepowQuad} is feasibly solved for $0 \leq p_{j,c} \leq 1$ to obtain the optimal power allocation coefficients $p_{k,c}^\star, $ where $c \neq C$ given by \eqref{optPowerAlloc1} in the next page. Using $p_{j,c}^\star$, the optimal power allocation coefficient for user $i$ is calculated as $p_{i,c}^\star = 1 - p_{j,c}^\star$.
Finally, the optimal time-slot allocation $t_c^\star$ substituting $p_{j,c}^\star$ in \eqref{time_pow_initial} can be obtained, and is given by \eqref{optTimeAlloc1} in the next page.
\begin{figure*}[b!]
\noindent\rule{\textwidth}{0.4pt}
\begin{equation}
\label{optPowerAlloc1}
    p_{j,c}^\star=\frac{- \sigma^2\left( |\mathbf{h}_i\mathbf{w}_c|^2 + |\mathbf{h}_j\mathbf{w}_c|^2\right) + \sigma\left(\sqrt{\sigma^2\left( |\mathbf{h}_i\mathbf{w}_c|^2 + |\mathbf{h}_j\mathbf{w}_c|^2\right)^2 + 4\,|\mathbf{h}_j\mathbf{w}_c|^2\, |\mathbf{h}_i\mathbf{w}_c|^4}\right)}{2\,|\mathbf{h}_i\mathbf{w}_c|^2|\, |\mathbf{h}_j\mathbf{w}_c|^2}.
\end{equation}
\begin{equation}
\label{optTimeAlloc1}
t_c^\star=
\begin{cases}
 \cfrac{R^\text{min}_{k,c}}{\log_2 \mathlarger{\left(1 + \frac{-\sigma\left( |\mathbf{h}_i\mathbf{w}_c|^2 + |\mathbf{h}_j\mathbf{w}_c|^2\right) + \sqrt{\sigma^2\left( |\mathbf{h}_i\mathbf{w}_c|^2 + |\mathbf{h}_j\mathbf{w}_c|^2\right)^2 + 4\,|\mathbf{h}_j\mathbf{w}_c|^2\, |\mathbf{h}_i\mathbf{w}_c|^4} }{2\,\sigma\,|\mathbf{h}_i\mathbf{w}_c|^2}\right)}}\;\;\text{if}\quad{c \,\in\, \{\mathbb{C} \setminus C\}}\\
 1 -{\sum\limits_{c \,\in\, \{\mathbb{C} \setminus C\}}{t_c}}\,\qquad \qquad \qquad \qquad \qquad \qquad \qquad\qquad\qquad\qquad\qquad \qquad\qquad\qquad\qquad\quad\;\text{if}\quad c=C.
\end{cases}
\end{equation}
\end{figure*}

\case{Paired users having different QoS requirements ($\mathlarger{R^\text{min}_{i,c} \neq  R^\text{min}_{j,c}}$)}
{ Finding the closed form solution analytically is dependent on the $\mathlarger{\nicefrac{R^\text{min}_{i,c}}{R^\text{min}_{j,c}}}$, by which the computational complexity increases significantly for small changes in given ratio. Thus, for the considered case, it is feasible to find joint power and time-slot allocation by solving \eqref{time_pow_initial} numerically rather than analytically satisfying the constraints \eqref{opt5} and \eqref{opt6}.
}
\end{caseof}
Now we derive the expression for optimal power allocation $p_{k,C}^\star$ for cluster $C$ with the highest sum of user channel strength. For the considered case of user $i$ having lower decoding order than user $j$ such that $\vartheta(i) < \vartheta(j) $, the optimal sum-rate maximization approach is to serve the user $i$ with power coefficient satisfying its QoS requirement and allocate the remaining power coefficient to user $j$. The derivation to obtain $p_{k,C}^\star$ based on this approach is as follows.

For the two-user case, since the IntraC at user $i$ is given by $\mathcal{I}(i,c) = \, 1 - p_{i,c} $, the rate of user $i$ can be written as 
\begin{gather*}
\gamma^c_{i\rightarrow i} = \frac{|\mathbf{h}_i \mathbf{w}_c|^2\; p_{i,c}} { {|\mathbf{h}_i \mathbf{w}_c|^2 \,(1 -  p_{i,c})} + \,\sigma^2}, \\
\gamma^c_{i\rightarrow i}\,|\mathbf{h}_i \mathbf{w}_c|^2 - \gamma^c_{i\rightarrow i}(p_{i,c})\,|\mathbf{h}_i \mathbf{w}_c|^2  + \gamma^c_{i\rightarrow i}\,\sigma^2 \,= \, |\mathbf{h}_i \mathbf{w}_c|^2\; p_{i,c}, \notag\\
\gamma^c_{i\rightarrow i}\,|\mathbf{h}_i \mathbf{w}_c|^2\, +\,\gamma^c_{i\rightarrow i}\,\sigma^2 \, = \, |\mathbf{h}_i \mathbf{w}_c|^2\; p_{i,c} + \gamma^c_{i\rightarrow i}(p_{i,c})\,|\mathbf{h}_i \mathbf{w}_c|^2,  \notag\\
\gamma^c_{i\rightarrow i}\,|\mathbf{h}_i \mathbf{w}_c|^2\, +\,\gamma^c_{i\rightarrow i}\,\sigma^2 \, = \, |\mathbf{h}_i \mathbf{w}_c|^2 \, p_{i,c} (1 \, + \,\gamma^c_{i\rightarrow i} ), \notag\\ 
p_{i,c} \,=\,\frac{\gamma^c_{i \rightarrow i}}{(1\,+\, \gamma^c_{i \rightarrow i})} \, + \, \frac{\gamma^c_{i \rightarrow i} \, \sigma^2}{|\mathbf{h}_i \mathbf{w}_c|^2 \, (1 \, + \, \gamma^c_{i \rightarrow i})} , \notag\\
p_{i,c}  \, = \, \frac{\gamma^c_{i \rightarrow i}}{(1\, + \, \gamma^c_{i \rightarrow i})} \,\left(1\, + \, \frac{\sigma^2}{|\mathbf{h}_i \mathbf{w}_c|^2}\right) \notag.
\end{gather*} 

Similarly, for user $k$ having decoding order $\vartheta(k)$, power allocation coefficient $p_{k,c}$ can be generalized as
\begin{align*}
\label{PfaValue}
p_{k,c}  \, = \, \frac{\gamma^c_{k \rightarrow k}}{(1\, + \, \gamma^c_{k \rightarrow k})} \,\left(1\, + \, \frac{\sigma^2}{|\mathbf{h}_k \mathbf{w}_c|^2} \, - \, \sum\limits_{i =  1}^{k  -  1} \,p_{i,c}  \right),
\end{align*}
where it is supposed $\gamma^c_{k \rightarrow k} = 2^{\nicefrac{R^{\text{min}}_{k,c}}{t_c}} - 1$ with $t_c$ being the allocated time-slot. Therefore, the generalized solution to optimal power allocation coefficient is given by \eqref{optPowerAlloc2}.

\begin{figure*}[b!]
\begin{equation}
\label{optPowerAlloc2}
p_{k,C}^\star = \, 
\begin{cases}
\, \left(1 - \dfrac{1}{2^{\nicefrac{R^{min}_{k,c}}{t_c}}}\right) \,\left(1\, + \, \dfrac{\sigma^2}{|\mathbf{h}_k \mathbf{w}_c|^2} \, - \, \sum\limits_{i =  1}^{k  -  1} \,p_{i,c}  \right) \,\text{if} \quad \vartheta(K_c) \geq \vartheta(k)\\
1 - p_{k,c} \qquad\qquad\qquad\qquad\qquad\qquad\qquad\quad\quad\;\text{if}\quad k=K_c.
\end{cases}
\end{equation}
\end{figure*}

Overall, the solutions to the user pairing and decoding order assignment as well as the active, passive beamforming and time and power allocation are summarized as a unified framework based on alternative optimization in $\textbf{Algorithm  \ref{Alg:PropAlg}}$. To facilitate faster convergence of the proposed algorithm, active and passive beamformers can be feasibly initialized using the estimated CSI. For example, the set of initial active beamforming vectors can be found using the maximal-ratio-transmission (MRT) precoding corresponding to the strongest user of the cluster given by
\begin{equation}
\label{feasInit1}
\mathbf{w}^{(0)}_{c} = \frac{{\mathbf{h}^{(0)}_{s,c}}}{\|{\mathbf{h}^{(0)}_{s,c}\|}} , \text{ where } {\mathbf{h}^{(0)}_{s,c}} = \argmax_{\mathbf{h}_{k,c} \in \{\mathbf{h}_{k,c}^{(0)}\}} \|\mathbf{h}_{k,c}\|^2.
\end{equation}
Similarly, these intial active beamformers can be used to feasibly initialize passive beamforming vectors given by
\begin{equation}
\label{feasInit2}
{\mathbf{u}^{\chi}_{c}}^{(0)} = \frac{ {\beta_{m}^\chi}^{(0)} \left(\text{ diag}\left(\mathbf{g}^H_{k,c}\right)\mathbf{H} \mathbf{w}^{(0)}_{c}\right)^H}{\text{ diag}\left(\mathbf{g}^H_{k,c}\right)\mathbf{H} \mathbf{w}^{(0)}_{c}},
\end{equation}where ${\beta_{m}^\chi}^{(0)}, p_{k,c}^{(0)}, \text{ and } t_{k,c}^{(0)}$ can be initialized randomly satisfying the unit-bound constraint.

\begin{lemma}
For a cluster c with distinct users $i,\,j \, \in \mathbb{K}_c$, given the active beamforming vector $\mathbf{w}_c$, passive beamforming vector $\mathbf{u}^\chi_c$, power allocation coefficients $\{p_{k,c}\}$ and time allocation $t_c$ optimized through \eqref{ABF}, \eqref{pbfdiscrete2}, \eqref{optPowerAlloc1}, \eqref{optPowerAlloc2} and \eqref{optTimeAlloc1}, iff the decoding order through $\textbf{Algorithm  \ref{Alg:UserPair}}$ is assigned such that $\vartheta(i) < \vartheta(j)$, then the following NOMA-SIC constraint is guaranteed
$$ R_{j\to i} \geq R_{i\to i}.$$
Thus, without significant loss of generality, \eqref{opt3} can be satisfied.
\end{lemma}

\begin{IEEEproof}
From \eqref{DecodingSINR} and \eqref{DecodingRate}, the decoding rates $R_{j\to i}$ and $ R_{i\to i}$ are given, respectively, as
\begin{subequations}
\begin{align*}
R_{j\to i} = t_c\left(\log_2 \left(1 + \frac{|\mathbf{h}_j \mathbf{w}_c|^2\; p_{i,c}} {{|\mathbf{h}_j \mathbf{w}_c|^2 \, p_{j,c}} + \,\sigma^2} \right)\right),\\
R_{i\to i} = t_c\left(\log_2 \left(1 + \frac{|\mathbf{h}_i \mathbf{w}_c|^2\; p_{i,c}} {{|\mathbf{h}_i \mathbf{w}_c|^2 \, p_{j,c}} + \,\sigma^2} \right)\right).
\end{align*}
\end{subequations}
In the above expressions, the decoding rates are functions of terms $|\mathbf{h}_j \mathbf{w}_c|^2$ and $|\mathbf{h}_i \mathbf{w}_c|^2$ only as the terms $p_{i,c}\,, p_{j,c},\text{and } \sigma^2$ are constants.
In addition, the active beamformer $\mathbf{w}_c$ in accordance with the given decoding order has been obtained such that the following inequality always holds true.
\begin{equation}
\label{LemmaIneq}
|\mathbf{h}_j \mathbf{w}_c|^2 \geq  |\mathbf{h}_i \mathbf{w}_c|^2.
\end{equation}  
Now, we write decoding rates as
\begin{subequations}
\begin{align*}
R_{j\to i} = t_c\left(\log_2 \left(1 + \frac{p_{i,c}} {{p_{j,c}} + \, \frac{\sigma^2}{|\mathbf{h}_j \mathbf{w}_c|^2 }} \right)\right),\\
R_{i\to i} = t_c\left(\log_2 \left(1 + \frac{p_{i,c}} {{p_{j,c}} + \, \frac{\sigma^2}{|\mathbf{h}_i \mathbf{w}_c|^2 }} \right)\right).
\end{align*}
\end{subequations}
From \eqref{LemmaIneq}, we can easily deduce that
$$R_{j\to i}  \geq R_{i\to i}\,.$$ 
\end{IEEEproof}

\begin{prop}
For a cluster c with users $i,\,j \, \in \mathbb{K}_c$ such that $\vartheta(i) < \vartheta(j)$, the active beamforming vector $\mathbf{w}_c$, passive beamforming vector $\mathbf{u}^\chi_c$, power allocation coefficients $\{p_{k,c}\}$ and time allocation $t_c$ optimized through \eqref{ABF}, \eqref{pbfdiscrete2}, \eqref{optPowerAlloc1}, \eqref{optPowerAlloc2} and \eqref{optTimeAlloc1} are guaranteed to ensure QoS such that
$$R_{k \to k} \geq R^{\text{min}}_{k,c} \quad k \in \{i,j\}.$$
Using $Lemma\,1$, we also obtain that
$$R_{j\to i}  \geq R_{i\to i} \geq R^{\text{min}}_{i,c}\,$$
$$ \Rightarrow R_{j\to i} \geq R^{\text{min}}_{i,c}, $$ which without loss of generality satisfies \eqref{opt4}.
\end{prop}

The proposed solutions to the aforementioned problems have polynomial-time complexity, and the individual computational complexity of the problem in \eqref{ABF} and \eqref{pbfsubp2} are $O(CK_c(2N_t^3 + 2K_c))$ and $O(CK_c(2M^3 + 2K_c))$, respectively, while the solution to \eqref{opt5} and \eqref{opt6} can be obtained analytically or numerically immediately.
\begin{algorithm}[t!]
 \caption{Proposed Algorithm for Sum-rate Maximization}
 \label{Alg:PropAlg}
\vspace{1mm}
 \begin{algorithmic}[1]
 \State \textbf{Input}: $\{\mathbf{h}_{k,c}\}, K , N_t , M, P_{max}, R_{k,c}^\text{min}, \sigma^2 ,\,$ threshold $\xi\,, \text{ and maximum iterations } \mathcal{T}_{max},\forall\,k \in \mathbb{K}_c\,, c \in \mathbb{C}$.
 \State Initialize with random user pairing to obtain $\{\mathbf{h}^{(0)}_{k,c}\} ,\,\forall\,k \in \mathbb{K}_c\,, c \in \mathbb{C}$
  \State Perform user pairing and decoding order allocation using \textbf{Algorithm \ref{Alg:UserPair}}$\left(K,\{\mathbf{h}_k\}\right)$ to obtain $\mathbb{K}_c^\star\,,\, {\vartheta({k_c})}^\star$.
  \State Initialize the optimization parameters: $\{\mathbf{w}^{(0)}_{c}\},$ $\{{\mathbf{u}^{\chi}_{c}}^{(0)}\}$,$\{p^{(0)}_{k,c}\}$,$\{t_{c}^{(0)}\}$ and threshold ${\xi}$ feasibly (e.g., using \eqref{feasInit1} and \eqref{feasInit2}).
  \Repeat
  \State Update $\mathbf{w}_c^{(i)}$ using ${\mathbf{u}_c^\chi}^{(i-1)}\,, p^{(i-1)}_{k,c}\,,t_{c}^{(i-1)} $ by solving the problem in \eqref{ABF};
  \State Calculate ${\mathbf{u}_c^\chi}^{(i)}$ using $\mathbf{w}_c^{(i)}\,, p^{(i-1)}_{k,c}\,,t_{c}^{(i-1)}$ by solving the problem in \eqref{pbfsubp2};
  \State Update ${\mathbf{u}_c^\chi}^{(i)}$ using \eqref{pbfdiscrete1} and \eqref{pbfdiscrete2};
  \State Update $\mathbf{h}^{(i)}_{k,c} \gets \mathbf{ g}_{k,c}^H {\mathbf{\Phi}^\chi_c}' \mathbf{H}\,, $ where ${\mathbf{\Phi}^\chi_c}' = \text{diag} ({\mathbf{u}_c^\chi}^{(i)})$;
  \State Update $p^{(i-1)}_{k,c}$ and $t_{c}^{(i-1)}$ via joint optimization using $\mathbf{h}^{(i)}_{k,c}\,,\mathbf{w}_c^{(i)}$ through \eqref{optPowerAlloc1}, \eqref{optPowerAlloc2} and \eqref{optTimeAlloc1};
  \If  {$ |R_{\text{sum}}^{(i)} - R_{\text{sum}}^{(i-1)}|^2 \leq \xi \; or\; i > \mathcal{T}_{max}$}
  \State$\{\mathbf{w}^{\star}_{c}\}  =  \{\mathbf{w}^{(i)}_{c}\},\,
   \{{\mathbf{u}^{\chi}_{c}}^{\star}\} =\{{\mathbf{u}^{\chi}_{c}}^{(i)}\} ,\,
   \{p^{\star}_{k,c}\} = \{p^{(i)}_{k,c}\} ,\,
  \{t_{c}^{\star}\} = \{t_{c}^{(i)}\},\,R^\star_{\text{sum}} = R^{(i)}_{\text{sum}}$;
  \State \textbf{break}
  \EndIf
  \State Set $i \gets i + 1$;
  \Until the value of the objective of the problem in \eqref{ProbForm} converges
  \State \textbf{Output}: $R^\star_{\text{sum}} ,\{\mathbf{w}^\star_{c}\},\{{\mathbf{u}^{\chi}_{c}}^\star\},\{p^\star_{k,c}\},\{t_{c}^\star\}$.
 \end{algorithmic} 
  \vspace{1mm}
 \end{algorithm}

\section{Simulation Results}
In this section, we present simulation results to evaluate the STAR-RIS mmWave communication system and validate the effectiveness of the proposed sum-rate maximization algorithm. We assume that the BS is located in a 3D space at $(x,y,z)=(0\text{ m},0 \text{ m}, 20 \text{ m})$ and communicating at 28 GHz frequency with the STAR-RIS placed at $(45\text{ m},-22\text{ m},0\text{ m})$ to serve users randomly scattered around it (i.e., within a circular region of radius $50 \text{ m}$ centered at the STAR-RIS). The parameters used for the simulation are given in Table \ref{tab:Simulation Parameters}.

\begin{table}[t!]
	\caption{Simulation Parameters}
	\label{tab:Simulation Parameters}
\centering
\renewcommand{\arraystretch}{1.1}
\resizebox{.48\textwidth }{!}{
\begin{tabular}{c c c}
\toprule
{\bf Notation} &{\bf Description}                                                                                                                                                                                                                           & {\bf Value} \\ \midrule
$P_{max}$ & BS transmit power &  30 dBm\\
$M$ &  Number of STAR-RIS elements  &  16  \\
$N$ & Number of BS antenna elements  &  16 \\
$K$ & Number of users & 6\\
$C$ & Number of clusters & 3\\
$PL_{do}$ & Path loss at reference distance & $60$ dBm \\
$\eta_{\text{(BS-STAR-RIS)}}$ &
Path-loss exponent of BS-STAR-RIS link &  $2.2$\\
$\eta_{\text{(STAR-RIS-User)}}$ &
Path-loss exponent of STAR-RIS-user link&  $2.8$\\
$\zeta$& Shadowing & $5.8$ dB  \\
$BW$ & Bandwidth & $10$ MHz  \\
$\sigma^2$ & Noise power & $-104$ dBm\\
$T_{\text{max}}$ & Channel coherence time & $650$ $\mu$sec\\
 \bottomrule
\end{tabular}}
\end{table}

%

\begin{figure}[t!]
\includegraphics[width=.5\textwidth]{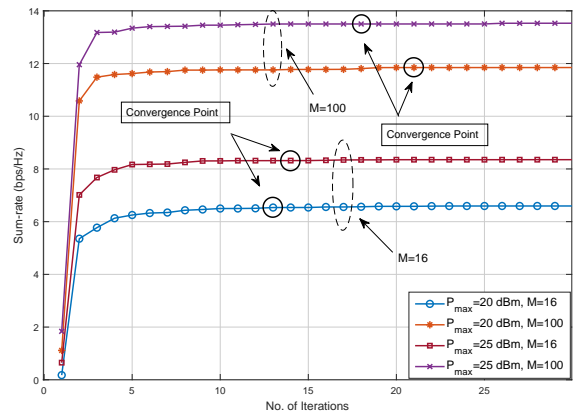}
\caption{Sum-rate vs. the number of iterations showing the convergence of \textbf{Algorithm \ref{Alg:PropAlg}}  }
\label{fig:SR_schemes}
\end{figure}
\begin{figure}[t!]
\includegraphics[width=.5\textwidth]{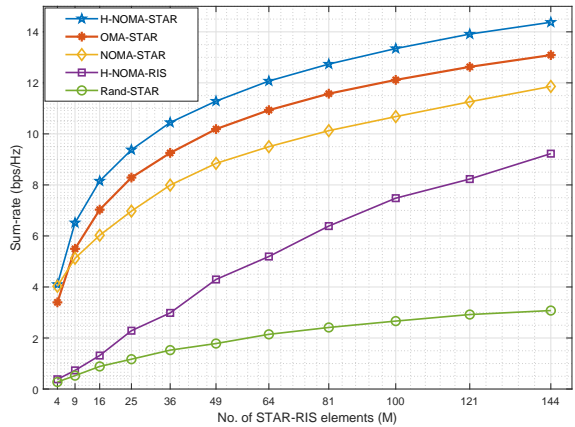}
\caption[width=.5\textwidth]{Sum-rate vs. the number of STAR-RIS elements $M$ for various schemes}
\label{fig:schemesComparison}
\end{figure}

Fig. \ref{fig:SR_schemes} shows the sum-rate optimized by the proposed \textbf{Algorithm \ref{Alg:PropAlg}} with the optimal user pairing for different values of $M$ and $P_{max}$. The results are obtained from Monte Carlo simulations over $10^4$ realizations using the maximum number of iterations $\mathcal{T}_{max}= 30$ and the convergence threshold $\xi = 0.1$ bps/Hz. First, it can be observed that we can achieve higher sum-rate, as $M$ and $P_{max}$ increase. It also shows that the proposed iterative algorithm converges for various values of parameters such as the STAR-RIS elements and BS power, i.e., $M \in\{16,\, 36\}$ and $P_{max} \in\{ 20,\,30\,\} $ dBm. As shown in the figure, the algorithm, initialized feasibly, converges to the optimal sum-rate value in less than 25 iterations even for low values of $\xi$. While power and time allocation is performed optimally in each iteration, the overall algorithm converges as the optimal active and passive beamformers $\mathbf{w}_{c}$ and $\mathbf{u}^\chi_{k,c}$ are achieved in an iterative fashion utilizing the CSI. The result also demonstrates that a larger number of STAR-RIS elements $M$ can be effectively used to obtain a higher sum-rate in an energy-constrained system because it is more likely to modify the blocked BS-STAR-RIS channels and improve the sum-rate by appropriately adjusting their amplitudes $\beta_m$ and phases $\theta_m$.

In Fig. \ref{fig:schemesComparison}, to better evaluate our proposed \textbf{Algorithm \ref{Alg:PropAlg}}, we compare it with four baseline schemes, i) STAR-RIS-assisted OMA-STAR, ii) STAR-RIS-assisted NOMA (NOMA-STAR), and iii) hybrid-NOMA-RIS (H-NOMA-RIS), iv) Rand-STAR, with different numbers of STAR-RIS elements $M$. For NOMA-STAR and H-NOMA-RIS, the users are served with random user pairing and decoding orders, whereas each user is served in its respective sub-time-slot in OMA-STAR. The time allocation for NOMA-STAR and H-NOMA-RIS is done using \eqref{optTimeAlloc1}, whereas for OMA-STAR the sub-time-slots allocation is made in a similar manner, as in \eqref{optTimeAlloc1}. For NOMA-STAR, we reproduce the results of \cite{refpaper_joint_star}, where each cluster is served simultaneously with its own beamformer, thus introducing inter-cluster-interference (InterCI) in the system model. Therefore, the optimal active and passive beamformers $\mathbf{w}_{c}$ $\mathbf{u}^\chi_{k,c}$ are realized by incorporating the InterCI as in (31) and (33) in \cite{refpaper_joint_star}. In the H-NOMA-RIS scheme, the STAR-RIS is set to operate only in reflection mode ($\beta^{\,r}_{m,c}\approx 1$), while the value of the transmission coefficient $\beta^{\,t}_{m,c}$ is negligible. An RIS is usually constructed as a metallic sheet, which has an intermediate copper back-plane to minimize the penetration of the impinged signals for maximum reflections \cite{IRSsheetplane}. In addition, a rectangular metallic sheet adds an average path loss of around $20$ dBm at mmWave (e.g., 28 GHz) \cite{IRSPL}, which is also accounted for in the STAR-RIS transmission users' channels $\mathbf{g}_{k,c}^t$. For OMA-STAR and H-NOMA-RIS, the beamformers are obtained using our proposed algorithm with CSI. The optimal power allocation $p_{k,c}$ is performed using \eqref{optPowerAlloc1} and \eqref{optPowerAlloc2} for STAR-NOMA and H-NOMA-RIS. For OMA-STAR, since the users are served in orthogonal sub-time-slots, there is no need for power allocation. For the Rand-STAR scheme, random user pairing, power allocation, active and passive beamforming are performed to compare the percentage increase in the sum-rate acquired. It can be observed from the figure that the achievable sum-rates of all the schemes (i.e., H-NOMA-STAR, OMA-STAR and NOMA-STAR) increase, as the number of STAR-RIS elements increases. This is due to the fact that the element array $\mathbf{u}_c^\chi$ will have more potential to form a pointed beam towards the intended user (in either reflection or transmission) by strengthening the respective end-to-end channel gain $\mathbf{h}_{k,c}$, as indicated by the Rand-STAR's increase in achievable sum-rate with increasing $M$.

Fig. \ref{fig:schemesComparison} also shows that the proposed scheme always outperforms other baseline schemes because operating STAR on H-NOMA not only leverages NOMA's effective resource allocation capability for higher SE, but also eliminates the InterCI to maximize the achievable sum-rate. For the H-NOMA-RIS scheme, as a result of the blocked BS-user (transmission) channel, highly unfair power allocation is essential to serve users, ensuring the QoS requirement $R^\text{min}_{k,c}$ particularly for lower values of $M$. The result shows that for the cluster's mean $R^\text{min}_{k,c} = 0.1$ bps/Hz, approximately $M > 64$ is required to serve all clusters satisfying the QoS constraint. While for a higher number of elements, i.e., $M > 64$, the majority of the sum-rate is achieved through reflection users via passive reflect beamformer $\mathbf{u}_{k,c}^r$ and the power allocation coefficients $p_{k,c}$ are adjusted to overcome RIS-blockage to serve transmission users. Therefore, to acquire a sum-rate equal to that of H-NOMA-STAR via H-NOMA-RIS, a very high value of $M$ or $P_{max}$ is required. For NOMA-STAR, the sum-rate is limited by the InterCI and hence, unless the users have uncorrelated channels, the proposed scheme yields greater sum-rate, whereas for OMA-STAR, further division of the resource (i.e., time-slots) reduces the sum-rate. Therefore, we can conclude that H-NOMA-STAR provides, in general, a flexible solution, whereas the other two schemes are highly subject to user-channel correlations and the resultant InterCI.

\begin{figure}[t!]
\includegraphics[width=.5\textwidth]{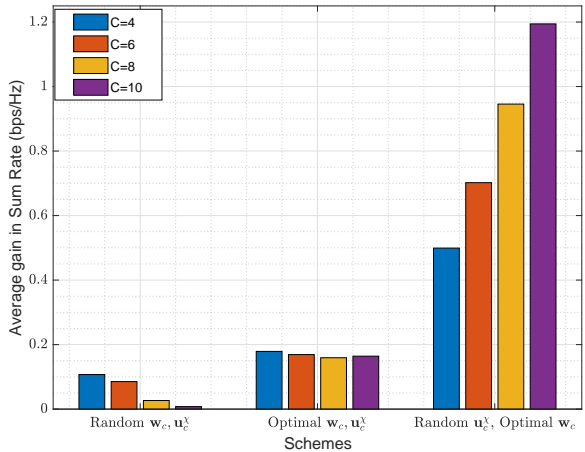}
\caption{The average gain in sum-rate obtained via pairing and decoding order with \textbf{Algorithm \ref{Alg:UserPair}}  }
\label{fig:SRvC}
\end{figure}

\begin{figure}[t!]
\includegraphics[width=.5\textwidth]{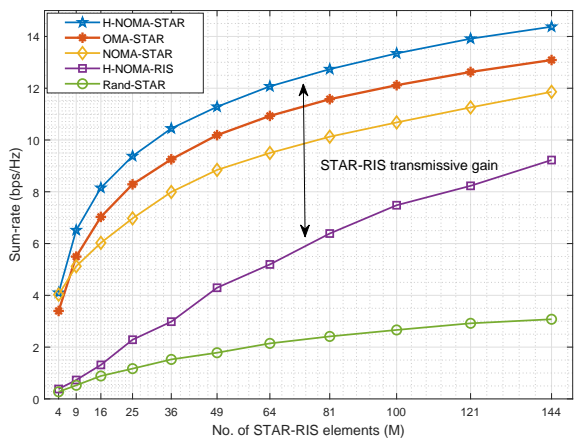}
\caption{The contribution of $\mathbf{w}_c,\,\mathbf{u}^\chi_{c}\,t_c\,,p_{k,c}$ to the sum-rat improvement with varying $M$, when $P_{max} = 25$ dBm, and mean $R^\text{min}_{k,c} = 0.1$ bps/Hz.}
\label{fig:SRvM_contrib}
\end{figure}

Fig. \ref{fig:SRvC} illustrates the average increase in the sum-rate obtained through the user pairing and decoding order strategy proposed in \textbf{Algorithm \ref{Alg:UserPair}}. The results are obtained with $M = 16$, $P_{max} = 30$ dBm, and mean cluster $R^\text{min}_{k,c} = 0.1$ bps/Hz by finding the optimal H-NOMA resource allocation parameters (i.e., $t_c\,,p_{k,c}$), and averaging the gain achieved with respect to random user pairing and decoding order over $10^4$ realizations. The result further presents three different cases proving that the gain achieved also depends on the number of clusters being served and the active and passive beamforming vectors:
\begin{itemize}
    \item \textbf{Random Active and Passive Beamformers}: The average gain obtained through optimal pairing and decoding order allocation for random $\mathbf{w}_c,\,\mathbf{u}^\chi_{c}$ is the lowest among the three cases because of the weak channel (non-optimized) correlations ${|\mathbf{h}_j \mathbf{h}_i^H|^2}$. It can also be observed that on a larger number of clusters $C$, serving users while guaranteeing their QoS  would require allocating smaller time slots, and hence, despite optimal pairing and decoding order, relatively smaller time-slot allocated to the clusters limits the achievable sum-rate gain. 
    \item \textbf{Optimized Active and Passive Beamformers}: In this case, since the optimal active and passive beamforming vectors $\mathbf{w}_c,\,\mathbf{u}^\chi_{c}$ are used to serve each cluster and its users, except for the few highly blocked users with weaker channel strength $\|\mathbf{h}_{k,c}\|$, each user can act as the sum-rate maximizer mainly depending on their STAR-RIS-user channels $\mathbf{g}^H_{k,c}$, therefore, the proposed channel correlation-based pairing, on average, yields a constant gain in the sum-rate. For the given simulation parameters, a gain of around $0.2\, \sfrac{bps}{Hz}$ for $C = 4$, $6$, $8$, $10$ is achieved.
    \item \textbf{Random Passive and Optimized Active Beamformer}: In this case, relatively more significant sum-rate gain is obtainable. This is due to the fact that obtaining optimal $\mathbf{w}_c$ can help increase a cluster's sum-rate even with random $\mathbf{u}^\chi_{c}$ (as highlighted in Fig. \ref{fig:SRvM_contrib}) by allocating longer time-slots to the cluster with the highest intrinsic sum-rate. Additionally, since the passive beamformers are not optimized, user-channel correlations ${|\mathbf{h}_j \mathbf{h}_i^H|^2}$ with different pairing and decoding order combination can have a noticeable difference in the sum-rate. Therefore, the optimal pairing and decoding order combination can be leveraged to have a decent increase in the sum rate, specifically at larger values of $C$ due to the increase in the number of possible pairing combinations.
\end{itemize}

\begin{figure}[t!]
\includegraphics[width=.5\textwidth]{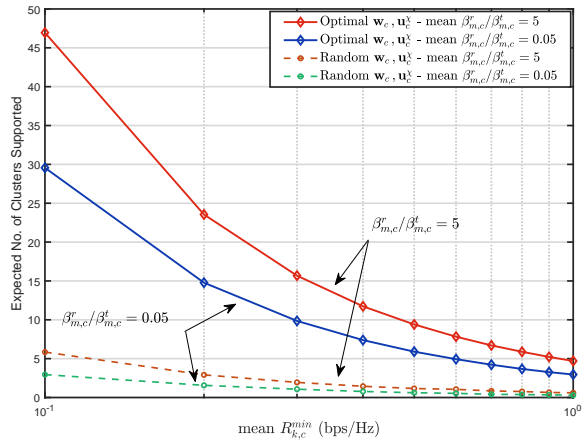}
\caption{The average number of clusters $C$ supported with $P_{max} = 30$ dBm, $M = 16$ for different mean ES-ratio. $\mathlarger{\nicefrac{\beta^r_{m,c}}{\beta^t_{m,c}}}$}
\label{fig:CvRmin}
\end{figure}

Fig. \ref{fig:CvRmin} depicts the expected number of clusters $C$ that can be served by the BS-STAR-RIS system for a simulation setting with $P_{max} = 30$ dBm, $M = 16$, such that all users are ensured their QoS requirements. We show that for various values of average $R_{k,c}^\text{min}$, the expected number of clusters also largely depends upon $\mathbf{w}_c,\, \mathbf{u}_{c}^\chi$ and the operating STAR-RIS ES-ratio $\nicefrac{\beta^r_{m,c}}{\beta^t_{m,c}}$. For analysis purposes, the results have been simulated by allocating higher decoding order to every user $i$ served via the STAR-RIS reflection such that $\vartheta(i) > \vartheta(j)$, where $\boldsymbol{state}(i) = r^i \,\text{and } \boldsymbol{state}(j) = t^j $, and $t_c$ and $p_{k,c}$ have been optimized using \eqref{optPowerAlloc1} and \eqref{optPowerAlloc2}. We also observe that the system with random $\mathbf{w}_c,\, \mathbf{u}_{c}^\chi$ is able to support, on average, no more than six clusters for $\nicefrac{\beta^r_{m,c}}{\beta^t_{m,c}} = 0.05\,,5$ , and the number of the mean clusters supported obviously decreases with increasing average $R_{k,c}^\text{min}$ because of the increase in the average time-slot $t_c$ allocated to a cluster. However, the optimal values of $\mathbf{w}_c,\, \mathbf{u}_{c}^\chi$ greatly reduce $t_c$, while ensuring the same QoS requirements. Consequently, the spare time-slot could potentially be used to serve a relatively larger number of clusters by the system on a given $R_{k,c}^\text{min}$. Moreover, by optimizing $\nicefrac{\beta^r_{m,c}}{\beta^t_{m,c}}$ to achieve a higher sum-rate (e.g. $\nicefrac{\beta^r_{m,c}}{\beta^t_{m,c}} = 5$, in this case), the average number of clusters supported can be further increased.

In Fig. \ref{fig:SRvM_contrib}, we highlight the impact of each design parameter, i.e. active and passive beamformers $\mathbf{w}_c,\,\mathbf{u}^\chi_{c}$, allocated resources $t_c\,,p_{k,c}$, and user-pairing and decoding order on the sum-rate. It can be observed that for $P_{max} = 30$ dBm and $C=3$ although optimizing the active beamformer $\mathbf{w}_c$ can yield a decent gain in the sum-rate. However, with increasing M, it only increases with a rate similar to that of random $\mathbf{w}_c,\,\mathbf{u}_c$ because of the constrained $P_{max}$ and the unoptimized passive beamformer. While the impact of optimal $\mathbf{w}_c$ is greater compared to $\mathbf{u}_c^\chi$ for $M < 16$, a significant gain (about 30\%) can be achieved by optimizing $\mathbf{u}_c^\chi$ for higher values of $M$ even with random $\mathbf{w}_c$. The result also gives a strong insight that the design of optimal active and passive beamformers is the most crucial one in a STAR-RIS-assisted system for higher sum-rate performance gains, as by only optimizing $\mathbf{w}_c$ and $\mathbf{u}_c$, approximately $80 \%$ of the total achievable sum-rate (through a fully optimized system), on average, is obtainable. It is also worth noticing that H-NOMA using optimal $t_c,\, p_{k,c}$ can further enhance the sum-rate again verifying the effectiveness of H-NOMA scheme. As depicted more clearly in Fig. \ref{fig:SRvC}, it can be re-examined that through optimal user pairing and decoding order, only a constant gain enhancement (about 5-10\%) is achievable.

\begin{figure}[t!]
\includegraphics[width=.5\textwidth]{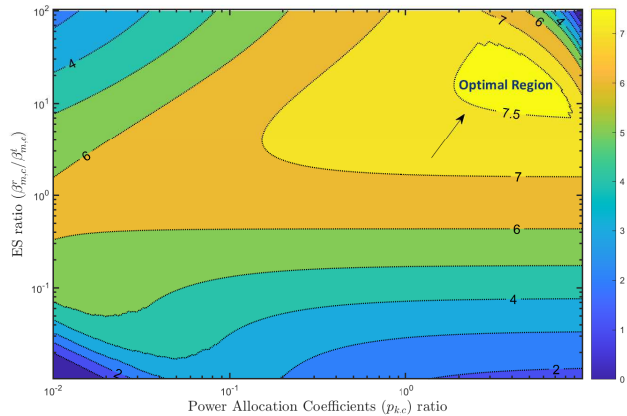}
\caption{Achievable sum-rate for different Energy Splitting ratio ($\nicefrac{\beta^r_{m,c}}{\beta^t_{m,c}}$) with $P_{max} = 25$ dBm, $ C=1\,,M =16$, mean $R^\text{min}_{k,c} = 0.1$ bps/Hz.}
\label{fig:SRvPAvES}
\end{figure}

The achievable sum-rate with optimal $\mathbf{w}_c\,, \mathbf{u}_c^\chi$ for various STAR-RIS ES ($\nicefrac{\beta^r_{m,c}}{\beta^t_{m,c}}$) and H-NOMA power allocation coefficients $p_{k,c}$ ratios is represented in Fig. \ref{fig:SRvPAvES}. To delve into the impacts of the ES and power allocation coefficients on the sum-rate, similar to Fig. \ref{fig:CvRmin}, we simulated two users $i$ and $j$ being served via STAR-RIS with decoding orders such that $\vartheta(i) > \vartheta(j)$, with $C=1$. It can be clearly seen from the contour map that for extreme values of ES i.e. $\nicefrac{\beta^r_{m,c}}{\beta^t_{m,c}} \gg 1,\, $and $\nicefrac{\beta^r_{m,c}}{\beta^t_{m,c}}\ll 1$, the sum-rate drops from the optimal value because despite obtaining optimal STAR-RIS phases $\theta^\chi_{m,c}$, the imbalance in $\nicefrac{\beta^r_{m,c}}{\beta^t_{m,c}}$ restricts the STAR-RIS QoS ensured coverage to the blocked (either reflection or transmission) side's users for the fixed $P_{max}= 25$ dBm. Likewise, allocating extreme power coefficients $p_{k,c}$ to the users also limits the sum-rate. For the given scenario, it can also be observed that because user $j$ has the lower decoding order, therefore, due to the IntraCI at user $j$ by user $i$, the optimal sum-rate can be achieved by allocating a relatively higher value of the power coefficient to user $i$ than $j$ such that $p_{i,c} > p_{j,c}$. It is also worth noticing that due to the IntraCI at $j$, since the reflection user $i$ has higher potential to increase the sum-rate, the rate of increase in the sum-rate due to slight increase in $\beta^r_{m,c}$ is higher than that caused by $\beta^t_{m,c}$. Hence, to operate in optimal region, the ES and power coefficients' ratio slightly favors the user with better channel condition, which is to be served with higher decoding order to obtain the optimal sum-rate.
\label{sSimRes}

\begin{figure}[t!]
\includegraphics[width=.5\textwidth]{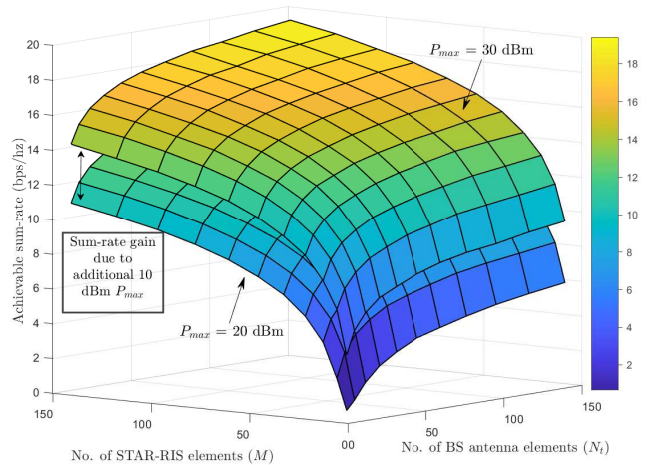}
\caption{Achievable sum-rate against different BS transmit antennas $N_t$ and STAR-RIS elements $M$ with mean $R^\text{min}_{k,c} = 0.1$ bps/Hz.}
\label{fig:SRvsMvsNT}
\end{figure}

Fig. \ref{fig:SRvsMvsNT} also examines the importance of two design parameters of the system model, i.e., the BS transmit antennas $N_t$ and the number of STAR-RIS elements $M$, on the achievable sum-rate using the proposed \textbf{Algorithm \ref{Alg:UserPair}} and \textbf{Algorithm \ref{Alg:PropAlg}}. For analysis purposes, the results are obtained on two different BS powers i.e., $P_{max} = 20$ dBm and $P_{max} = 30$ dBm. We observe that the achievable sum-rate is more sensitive to the STAR-RIS elements $M$ compared to the BS transmit antennas $N_t$. This is because on increasing $N_t$ (while keeping $M$ constant), the end-to-end BS-user channel $\mathbf{h}_{k,c}$ will cause the major increase in the sum-rate as the active beamforming vector $\mathbf{w}_c$ will have an approximately constant effect on the sum rate due to the maximum power constraint in \eqref{opt2}. On the other hand, increasing $M$ will cause more elements to coherently contribute to increase the sum-rate using their corresponding tunable coefficients $\beta^\chi_{m,c}$ and  independent phases $\theta_{m,c}^{\chi}$.

\section{Conclusion}
    
\label{sConclusion}
In this paper, we have investigated a STAR-RIS-assisted mmWave communication for sum-rate maximization and proposed a comprehensive optimization framework in the presence of blockages. We have handled the coupled non-convex optimization problem by solving sub-problems for the active and passive beamforming optimization. Moreover, we have also studied the appropriate multiple access scheme for the given scenario. Our results have confirmed that the proposed H-NOMA scheme outperforms the conventional NOMA and OMA. A channel correlation-based technique has been proposed to group a pair of reflective and transmissive users and find their NOMA decoding orders. Analytical solutions to the joint power-time allocation problem have also been derived for the optimal H-NOMA operation. Our results have shown that the STAR-RIS transmission property presents an ideal solution to the blocked communication coverage, since a specific number of STAR-RIS elements can be leveraged to obtain significant gains over the conventional RIS regardless of the chosen multiple access scheme. To fully exploit the advantages of the STAR-RIS-aided H-NOMA communication, more works on the accurate downlink CSI acquisition through STAR-RIS is required to realize the enhanced beamforming gains in practice. Moreover, as suggested earlier, the user pairing algorithm with different cluster sizes is also worth studying to further increase the sum-rate. Finally, although several works have explored the optimal deployments of the conventional RISs, the deployment strategies of STAR-RISs to improve the existing networks in the light of the stochastic geometry can be included in the future extensions.
\IEEEpeerreviewmaketitle

\bibliographystyle{IEEEtran}
\bibliography{refs}

\end{document}